\documentclass[letterpaper,11pt]{emulateapj}
\usepackage{epsfig}
\usepackage{natbib}
\usepackage{graphics}
\usepackage{graphicx}
\usepackage{multirow}
\usepackage{amssymb}
\usepackage{color}
\usepackage{boxedminipage}

\def \kms {{\rm km~s$^{-1}$}}
\def \kmsMpc {{\rm km~s$^{-1}$~Mpc$^{-1}$}}

\shorttitle{CF2 Data}
\shortauthors{Tully et al.}

\begin{document}

\title{Cosmicflows$-$2: The Data}

\author{R. Brent Tully$^{1}$}
\author{H\'el\`ene M. Courtois$^{1,2}$}
\author{Andrew E.  Dolphin$^3$}
\author{J. Richard Fisher$^4$}
\author{Philippe H\'eraudeau$^5$}
\author{Bradley A. Jacobs$^{1}$ }
\author{Igor D. Karachentsev$^6$}
\author{Dmitry Makarov$^6$}
\author{Lidia Makarova$^6$}
\author{Sofia Mitronova$^6$}
\author{Luca Rizzi$^7$}
\author{Edward J. Shaya$^8$}
\author{Jenny G. Sorce$^2$}
\author{Po-Feng Wu$^1$}
\affil{$^1$Institute for Astronomy, University of Hawaii, 2680 Woodlawn Drive, HI 96822, USA}
\affil{$^2$Universit\'e Claude Bernard Lyon I, Institut de Physique Nucleaire, Lyon, France} 
\affil{$^3$Raytheon Company, 1151 E Hermans Rd, Tucson, AZ, 85756, USA}
\affil{$^4$National Radio Astronomy Observatory, 520 Edgemont Road, Charlottesville, VA 22903, USA}
\affil{$^5$Argelander-Institut f\"ur Astronomie, Auf dem H\"ugel 71, D-53121, Bonn, Germany}
\affil{$^6$Special Astrophysical Observatory, N Arkhyz, KChR, 369167, Russia}
\affil{$^7$W.M. Keck Observatory, 65-1120 Mamalahoa Hwy, Waimea, HI 96743, USA}
\affil{$^8$Department of Astronomy, University of Maryland, College Park, MD 20742, USA}

\begin{abstract}
{\it Cosmicflows-2} is a compilation of distances and peculiar velocities for over 8000 galaxies.  Numerically the largest contributions come from the luminosity-linewidth correlation for spirals, the TFR, and the related Fundamental Plane relation for E/S0 systems, but over 1000 distances are contributed by methods that provide more accurate individual distances: Cepheid, Tip of the Red Giant Branch, Surface Brightness Fluctuation, SNIa, and several miscellaneous but accurate procedures.  Our collaboration is making important contributions to two of these inputs: Tip of the Red Giant Branch and TFR.  A large body of new distance material is presented.  In addition, an effort is made to assure that all the contributions, our own and those from the literature, are on the same scale.  Overall, the distances are found to be compatible with a Hubble Constant H$_0 = 74.4 \pm 3.0$~\kmsMpc.  The great interest going forward with this data set will be with velocity field studies.  {\it Cosmicflows-2} is characterized by a great density and high accuracy of distance measures locally, falling to sparse and coarse sampling extending to $z=0.1$.
\end{abstract}	


\section{Introduction}

This paper presents a compendium of galaxy distances and peculiar velocities that is being called {\it Cosmicflows-2}.  A precursor catalog by \citet{2008ApJ...676..184T} has retroactively been called {\it Cosmicflows-1}.  In both cases, the components of the catalogs are a mix of new material and information from the literature.  {\it Cosmicflows-1} provided, at the time, the densest coverage of distances locally but it was severely restricted by a cutoff of 3000~\kms.  The new {\it Cosmicflows-2} enhances the density of coverage locally and extends coverage sparsely to 30,000~\kms.

The current compilation draws on distance determinations by six distinct methods.  In four cases, all the base material is drawn from the literature and the endeavor here has been to assure a uniform scaling.  In two important cases, our collaboration has made major observational contributions, with much of our material being released for the first time with this publication.

The six independent methodologies have distinct merits and deficiencies.  The samples are now getting sufficiently large with all six procedures that each has good overlap with others.  Multiple measurements of individual galaxies and within groups and clusters are creating an increasingly tight lattice of distances on a common scale.

The zero point of the distance scale is set by two independent constructions that are shown to be in agreement.  The first derives from the Hubble Space Telescope (HST) distance scale key project \citep{2001ApJ...553...47F} with primary emphasis on the application of the Cepheid Period-Luminosity Relation (Cepheid PLR).  There have been recent modest refinements to this scale \citep{2011ApJ...730..119R, 2012ApJ...758...24F}.

Cepheid variables are young stars, frequently in regions of obscuration, and only present in galaxies currently forming stars.  Our second route to a zero point calibration strictly involves old stars.  Distances are determined from the luminosities  of Red Giant Branch stars at the onset of core Helium burning, at a location in a stellar color-magnitude diagram (CMD) known as the Tip of the Red Giant Branch (TRGB).  For this discussion, we avail of photometry obtained with HST with filters that approximate $I$ band.  There is now considerable experience that demonstrates the viability of TRGB measurements in this band \citep{1993ApJ...417..553L, 2006AJ....132.2729M, 2009ApJ...690..389M, 2009ApJS..183...67D}.  The zero point is provided by bootstrapping from observations of spheroidal companions to the Milky Way that link TRGB and Horizontal Branch magnitudes.  These latter are fixed to an absolute scale through studies of globular clusters.  \citet{2007ApJ...661..815R} demonstrated that this Population II path to a calibration gives extragalactic distances that agree with the Cepheid PLR scale at the level of 0.01 mag.  Further confirmation has come from the agreement in measured distances to NGC4258 from Maser observations, the Cepheid PLR, and TRGB \citep{1999Natur.400..539H, 2007ApJ...661..815R, 2008ApJ...689..721M, 2011ApJ...730..119R}.  Essentially all galaxies have an old population so are candidates for the TRGB methodology.   From experience, observations in a single orbit with HST result in a TRGB distance determination with 5\% accuracy for a galaxy within 10 Mpc.

The Surface Brightness Fluctuation (SBF) method \citep{2001ApJ...546..681T, 2010ApJ...724..657B} has a physical basis that is closely tied to TRGB.  Each is a characterization of stars at or near the Tip of the Red Giant Branch, in one case through resolution of individual stars and in the other case through statistical properties of unresolved populations.  Most SBF work to date has been carried out at ground based observatories which limits the depth of measurements to about 40 Mpc.  Considerably better can be done with HST but so far serious exploitation of this resource has been limited to studies of the Virgo and Fornax clusters \citep{2009ApJ...694..556B}.  SBF is only effective with systems dominated by old stars.

The fourth and fifth methods to be discussed are less accurate on an individual basis but can be applied to very large samples over a wide range of distances.  They share a common physical basis, linking velocity fields and luminosities through the separate relationships of these observables with mass.  One of these has come to be called the Fundamental Plane (FP) method \citep{1987ApJ...313...59D, 1987ApJ...313...42D} an extension of the Faber-Jackson Relation \citep{1976ApJ...204..668F} involving three parameters: velocity dispersion, luminosity, and a characteristic dimension (or surface brightness).  The FP is useful for the measurement of distances to early-type galaxies and is effective with studies of clusters where errors can be reduced by averaging over a number of targets.

The related method pertains to disk systems.  A correlation between rotation rate and luminosity is exploited \citep{1977A&A....54..661T}, giving a distance measurement relation hereafter referred to as TFR.  This methodology can be applied to normal spiral galaxies as long as they are not too face-on or in such close proximity to another system as to be confused or distorted.  Approximately 40\% of galaxies in an apparent magnitude limited sample can meet these criteria and they lie in the full range of environments hosting galaxies.  This method, then, is most important for providing a high density of observations over a wide range of distances and local conditions.  It will be discussed that we draw on two only partially independent sources of TFR measurements; one from the literature \citep{2007ApJS..172..599S} and one based on observations and processing by our team.

This overview is completed by mention of the sixth technology involving Type Ia supernovae (SNIa).  It has been amply demonstrated that SNIa events can be calibrated to acquire accurate distances \citep{1993ApJ...413L.105P, 2007ApJ...659..122J, 2009ApJ...700.1097H}.  The current limited deficiency with this otherwise splendid method is the low density of coverage because of the serendipitous nature of events.  It is a great benefit, though, that distances are available 
well out to the range where peculiar velocities are a negligible fraction of cosmic expansion velocities.

Sections 2 and 3 will discuss the original material that we bring to {\it Cosmicflows-2}, based respectively on the TRGB and TFR procedures.  Section 4 will summarize what has been done to assure consistent distances from the six separate procedures, mixing the new with the literature.  The composite catalog will be presented in the Section 5.  At that point there will be a discussion of galaxy groupings drawn from redshift catalogs because there will be distance and velocity averaging and luminosity integration over groups.  There follows a discussion of results in Section 6.

\section{Tip of the Red Giant Branch Distances}

\begin{figure}[h!]
\includegraphics[scale=0.76]{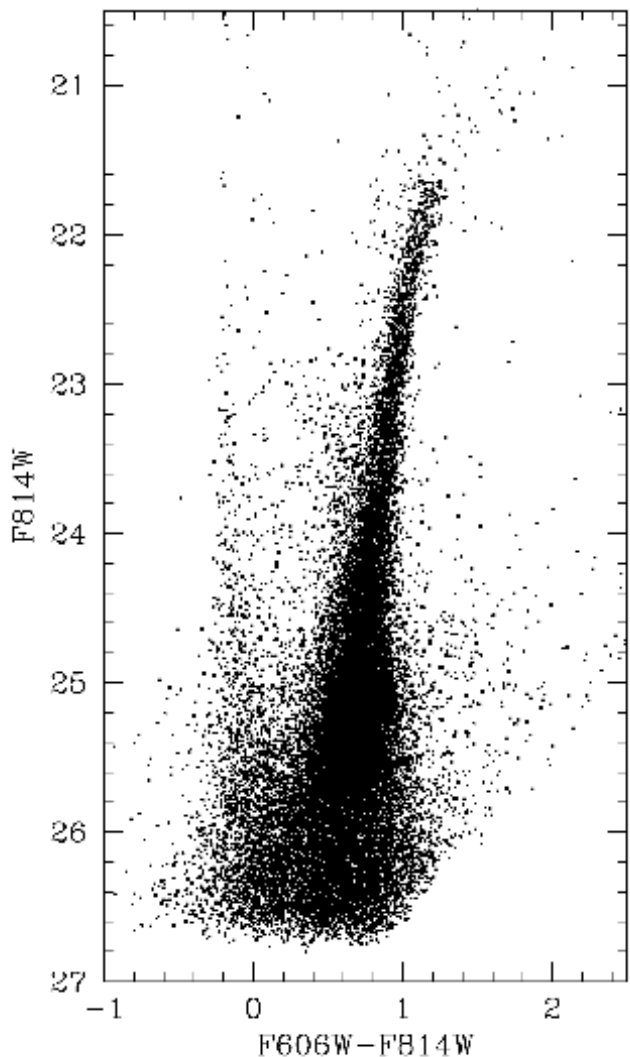}
\caption{Color-magnitude diagram for UGC4879.  The TRGB is fit at F814W=21.63, giving a distance of 1.37 Mpc.}
\label{cmd_u4879}
\end{figure}

The TRGB method for acquiring galaxy distances nicely complements the Cepheid PLR method.  The RGB arises in populations older than 1 Gyr while Cepheid variables are signatures of young populations.  Cepheids are bright but multi-epoch observations are required to discover and characterize them.  RGB stars are fainter but can be characterized with a single sequence of observations in two passbands (see Figure~\ref{cmd_u4879}).  Cepheids, being young, frequently suffer from local obscuration whereas RGB stars, being old and widely distributed, can be studied in minimally obscured places.  The properties of the RGB depend on metallicity and age but variations are minimal and well calibrated at the optical $I$ band \citep{2007ApJ...661..815R} especially at low metallicities and all observed galaxies have a low metallicity component in their halos.  With Hubble Space Telescope (HST) and the Advanced Camera for Surveys (ACS) a distance with an accuracy of 5\% can be obtained for galaxies within 10 Mpc with a single HST orbit.  This accuracy is comparable to what can be achieved with the Cepheid PLR method.  Currently roughly 3 dozen distances have been obtained with HST based on the Cepheid PLR.  By comparison, about 300 TRGB distances have been acquired with HST using a third as many orbits.

Given the tremendous gain in efficiency over the Cepheid PLR with comparable achievable accuracies, it can be argued that the TRGB method is the current gold standard for distance determinations to nearby galaxies.  Our collaboration has undertaken the task of gathering together all the useful observations ever taken with HST with either ACS or Wide Field/Planetary Camera~2 (WFPC2), whether from programs we initiated or from the archive, and analyzed this material in a homogenous way.

The basic requirements for an HST observation to be useful are adequately long integrations (half orbit or more each) in the F814W filters that approximate Cousins $I$ band and one other filter.  The TRGB is determined from F814W magnitudes.  The second filter provides color discrimination that isolates the RGB and provides the information needed to make the minor metallicity/age corrections.  Almost always, the second filter is F555W or F606W although F475W is used on rare occasions.  The more blueward filters give better stellar population discrimination in principle but for distance measurements with the TRGB method the F606W filter is strongly favored because blueward filters loose red stars, the ones we care about.   Even with the F555W filter there is the danger of seriously clipping the red edge of the RGB.  Usually there is not a problem.  A large fraction of the observations now come from programs associated with this collaboration, assuring an appropriate pairing of two band exposures and a large fraction of the remainder come from programs that employed multi-orbit observing strategies, generating products of stacked images with considerable sensitivity.

While variations in observing strategies between programs with a given HST camera can be reasonably accommodated, the differences in data quality between ACS and WFPC2 are radical.  It was mentioned that a single HST orbit provides a distance for a galaxy out to 10 Mpc using ACS but material of the same quality in a single orbit with WFPC2 is limited to 4.5 Mpc.  Our program was born with WFPC2 but came of age with ACS.

 The TRGB methodology has been modified and improved over the years.  Pioneers \citet{1990AJ....100..162D} and \citet{1993ApJ...417..553L} brought awareness of the utility of the TRGB. \citet{1996ApJ...461..713S} refined the use of the Sobel filter as a means of identifying the TRGB location in a Color-Magnitude Diagram (CMD).   \citet{2002AJ....124..213M} introduced a more sophisticated maximum likelihood procedure for locating the tip.  \citet{2006AJ....132.2729M} introduced modifications to the maximum likelihood analysis including recovery of artificial stars to monitor completeness and photometric errors.  It is this procedure that is used in the present study.
 
\begin{figure}[]
\centering
\includegraphics{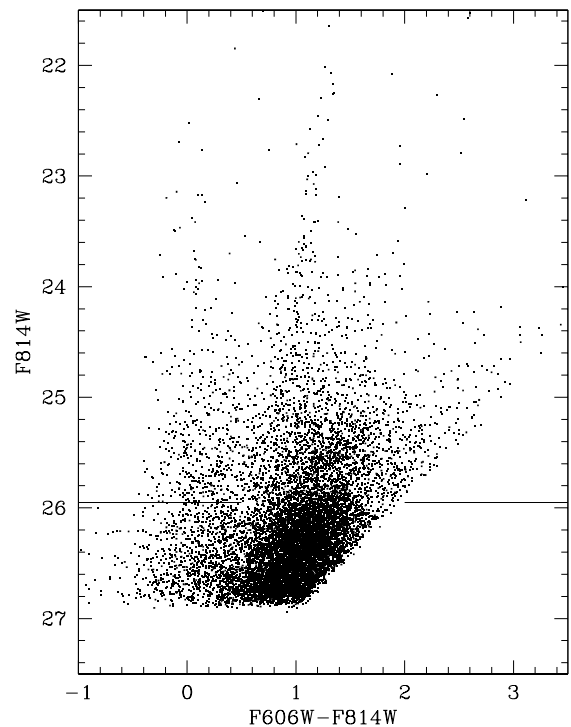}
\vspace{95mm}
\caption{Color-magnitude diagram for DDO52.  The TRGB is fit at F814W=25.95, giving a distance of 9.84 Mpc.}
\label{cmd_ddo52}
\end{figure}

 HST has made it practical to get distances for large samples of galaxies.  HST provides tenth arcsec resolution with stable point spread functions over sufficient (arcmin scale) fields.  From the ground it took a Herculean effort to address targets beyond the Local Group \citep{2002AJ....124..213M}.  Members of our collaboration were early users of WFPC2 on HST with the specific interest in measuring galaxy distances \citep{2002A&A...383..125K, 2002A&A...385...21K, 2002A&A...389..812K, 2003A&A...404...93K, 2003A&A...398..479K, 2003A&A...408..111K, 2003A&A...398..467K}.  With the availability of ACS the possibilities became more interesting.  A $S/N=5$ cutoff in the $I$ band equivalent F814W filter moves from 25.3 with WFPC2 to 27.0 with ACS in a single orbit observation in two colors. The TRGB should lie at least 1 mag brighter than this limit for a robust distance determination \citep{1995AJ....109.1645M, 2006AJ....132.2729M, 2009ApJ...690..389M}.   These conditions lead to the practical limits on distance modulus measures of $\sim 28$ with WFPC2 and $\sim 30$ with ACS.  Figure~\ref{cmd_ddo52} provides an example of a CMD acquired with ACS in one orbit of a galaxy at almost 10 Mpc.  Through programs managed by members of our collaboration the number of galaxies observed with ACS has grown quite large.  Other programs have tended to give detailed attention to modest samples \citep{2008ApJ...689..721M, 2009ApJS..183...67D}.  We presently have TRGB distances to 297 galaxies, with 197 of these coming from observations within this collaboration.
 
\begin{figure}[h!]
\includegraphics[scale=0.44]{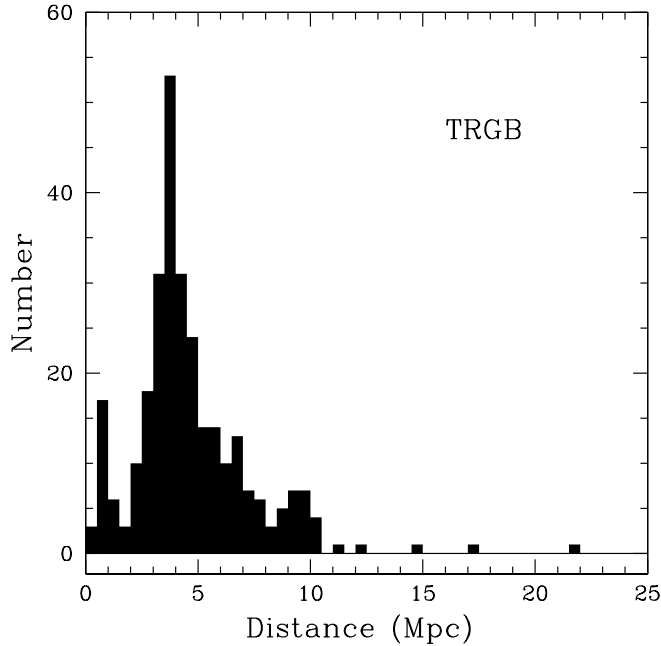}
\caption{Histogram of distances for almost 300 galaxies from HST TRGB observations.}
\label{histDtrgb}
\end{figure}

As mentioned, whatever the HST program source, the analysis is carried out with a consistent procedure.  The stellar photometry uses a program developed by one of us: HSTPHOT with WFPC2 images and the updated DOLPHOT with ACS images \citep{2000PASP..112.1383D}\footnote{http://americano.dolphinsim.com/dolphot/}.  The RGB tip determination follows the maximum likelihood procedures described by \citet{2006AJ....132.2729M}.  The zero point calibrations in alternative HST filter combinations were derived by \citet{2007ApJ...661..815R}.  The compilation of the data for public access in the Extragalactic Distance Database (EDD)\footnote{http://edd.ifa.hawaii.edu} is discussed in \citet{2009AJ....138..332J}.  See the catalog {\it CMDs/TRGB} in EDD for a tabulation of results and CMD and images for each individual galaxy.  Figure~\ref{histDtrgb} gives a histogram of the TRGB distances currently available.  The trickle of points beyond 10 Mpc come from programs with multi-orbit observations.

It has been demonstrated that there is good agreement between TRGB and Cepheid PLR distances \citep{2007ApJ...661..815R} and between these two and the Maser distance to NGC4258 \citep{2006ApJ...652.1133M, 2008ApJ...689..721M}.  The agreement will be re-evaluated with the discussion of the merging of alternative procedures in Section 4.

\section{Luminosity $-$ Linewidth Distances}

The correlation between the luminosity of a spiral galaxy and its rate of rotation \citep{1977A&A....54..661T}, the TFR,  can be used to determine distances with individual accuracies of 20\%.  Other methods provide greater precision but are not useful overly nearly such a wide range of distances and conditions.  {\it Cosmicflows-2} incorporates TFR distances from two alternative sources.  One source is the culmination of the important Cornell program in the data set called SFI++ \citep{2007ApJS..172..599S}.  The other source is a new assembly of material collected within this collaboration and presented here for the first time.  Discussions regarding SFI++ will be put off to Section 4 when we deal with the integration of alternative distance measures.  The current section is devoted to the work we have done to acquire TFR distances.

\begin{figure}[h!]
\includegraphics[scale=0.44]{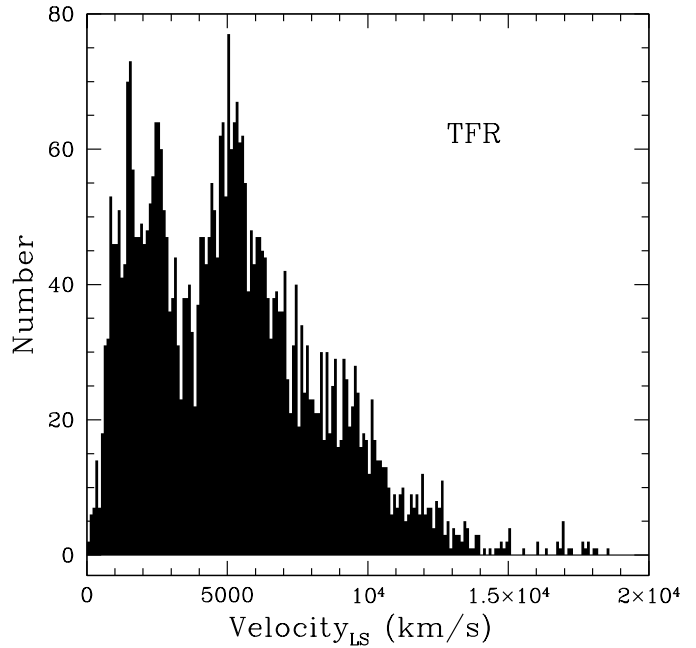}
\caption{Histogram of velocities (Local Sheet rest frame) for 4168 galaxies with TFR distances.}
\label{histVtf}
\end{figure}

Our previous major release of TFR distances was with the catalog now called {\it Cosmicflows-1} \citep{2008ApJ...676..184T} and followed procedures described by \citet{2000ApJ...533..744T}.  The current release involves two important advances.  The first is the scope of the sample.  Distance measures in {\it Cosmicflows-1} were limited to 3300~\kms\ while now TFR determinations are included for galaxies with redshifts almost as great as 20,000 \kms\ (see Figure~\ref{histVtf} for a histogram of the redshift dependence of our TFR measures).  The other important advance arises from a new definition of the rotation parameter.

\subsection{HI  Linewidths}

Details about how we derive a measure of maximum rotation velocities are provided by \citet{2009AJ....138.1938C, 2011MNRAS.414.2005C}.  We focus entirely on the interpretation of 21 cm HI profiles.  The data are provided by our observations with the Green Bank Telescope (GBT) or the Arecibo and Parkes telescopes or are acquired from archival sources related to the Arecibo Telescope, the National Radio Astronomy Observatory (NRAO) 140-foot and 300-foot telescopes and GBT, or the Nan\c cay, Parkes, and Effelsberg telescopes.  Each of these alternative sources gives us digital records of fluxes in wavelength bins.

With {\it Cosmicflows-1} and earlier we used an analog measure of an HI profile linewidth, $W_{20}$, the width of a profile at 20\% of peak intensity.  We now use a measure called $W_{m50}$, the width at 50\% of the mean flux per channel over the range of channels containing 90\% of the total flux.  The exclusion of 5\% of the flux at each of the high and low velocity extremes reduces ambiguity in the channel count due to profile wings.  The new linewidth definition is a more robust measure than $W_{20}$ with asymmetric or single peaked profiles. \citet{2005ApJS..160..149S} and \citet{2007AJ....134..334C} have discussed alternative linewidth characterizations.  Our choice is justified in \citet{2009AJ....138.1938C}.  We describe there the adjustments that are made to account for instrumental and redshift broadening
\begin{equation}
W^c_{m50} = {W_{m50} \over (1+z)} -2\Delta v \lambda
\label{Wc}
\end {equation}
where $z$ is redshift, $\Delta v$ is the smoothed spectral resoution, and $\lambda=0.25$ is an empirically determined constant.  Next we translate the observed linewidth parameter to a statistical measure of twice the maximum rotation speed
\begin{eqnarray}
\nonumber
W^2_{mx} = (W^c_{m50})^2 + (W_{t,m50})^2 [ 1-2e^{-(W^c_{m50}/W_{c,m50})^2}] \\ 
- 2W^c_{m50}W_{t,m50}[1-e^{(W^c_{m50}/W_{c,m50})^2}]
\label{WR}
\end{eqnarray}
with $W_{c,m50} = 100$~\kms\ describing the transition from boxcar (horned) profiles to roughly gaussian profiles and $W_{t,m50} = 9$~\kms describing the effects of thermal broadening.  Finally a de-projection translates from the observed inclination, $i$, to edge-on orientation in order to arrive at the desired parameter $W^i_{mx}$
\begin{equation}
W^i_{mx} = W_{mx} / {\rm sin}i .
\label{Wmx}
\end{equation}

The uncertainty in a linewidth measurement $W_{m50}$ is determined in the first instance by the ratio of the signal $S$ defined by the mean flux per channel in the spectral line to the noise $N$ in channels outside the spectral line

$e_W = 8$ \kms\ ~~~~~~~~~~~~~~~~~~~~~~~~~~~ if  $ S/N > 17$

$e_W = 21.6 - 0.8 S/N$ \kms\  ~~~~~~~~~~ if $ 2 < S/N < 17$

$e_W = 70 - 25 S/N$ \kms\ ~~~~~~~~~~~~~ if $ S/N < 2$.

In order to be considered useful for a TFR distance determination we require $e_W \le 20$~\kms.  After an initial automatic fit to line profiles from a computer algorithm, profiles and fits are inspected by eye.  Attention is given to possibilities of confusion from neighbors or to anomalies such as might arise from interactions.  The result of the inspection might be that a profile is rejected as a candidate for a distance determination, in which case $e_W$ is set greater than 20~\kms, or on rare occasions, that a profile with low $S/N$ is considered adequate, whence $e_W$ is set by hand to 20~\kms.

\begin{figure}[h!]
\includegraphics[scale=0.64, angle=270]{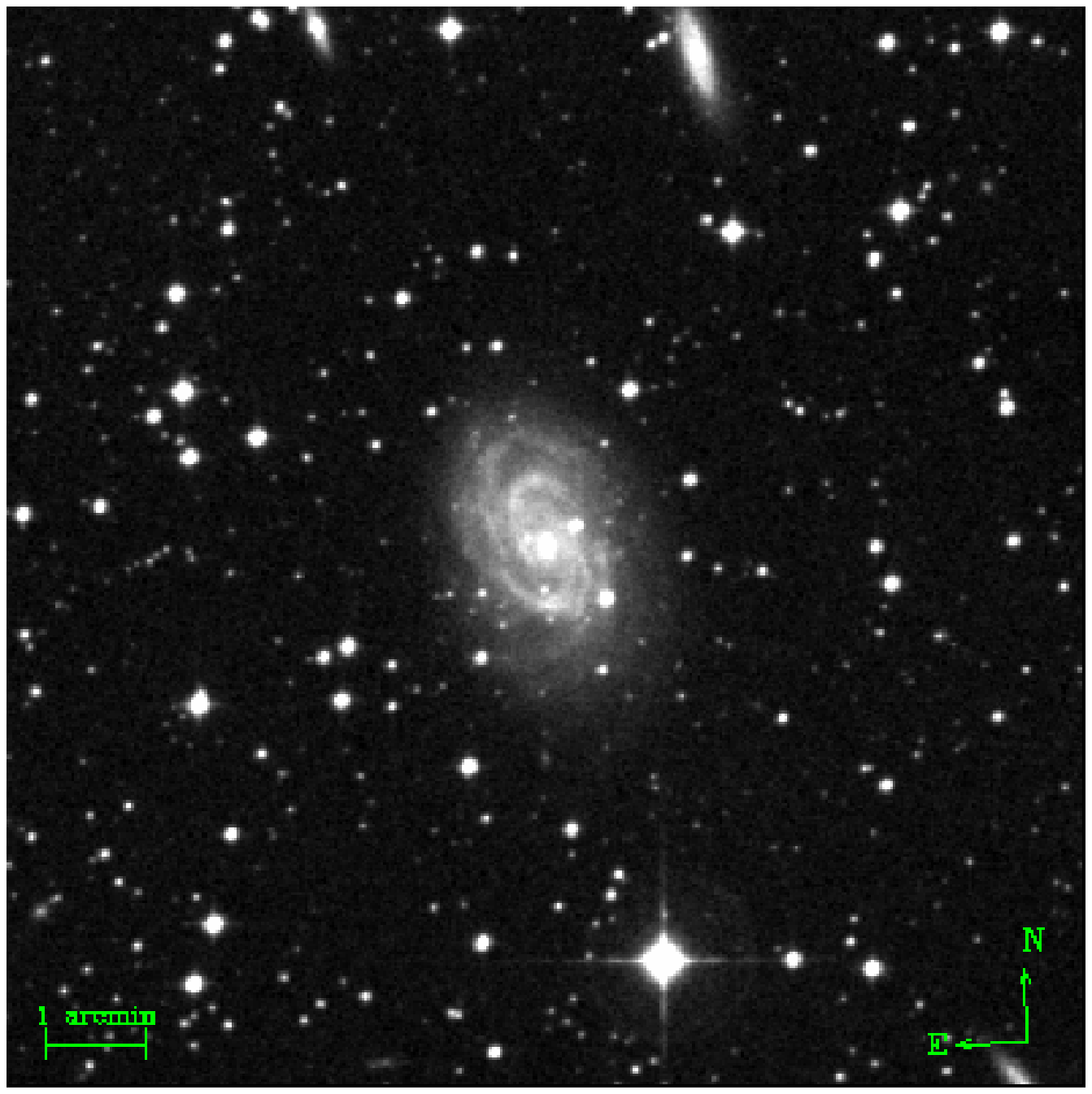}
\includegraphics[scale=0.52, angle=270]{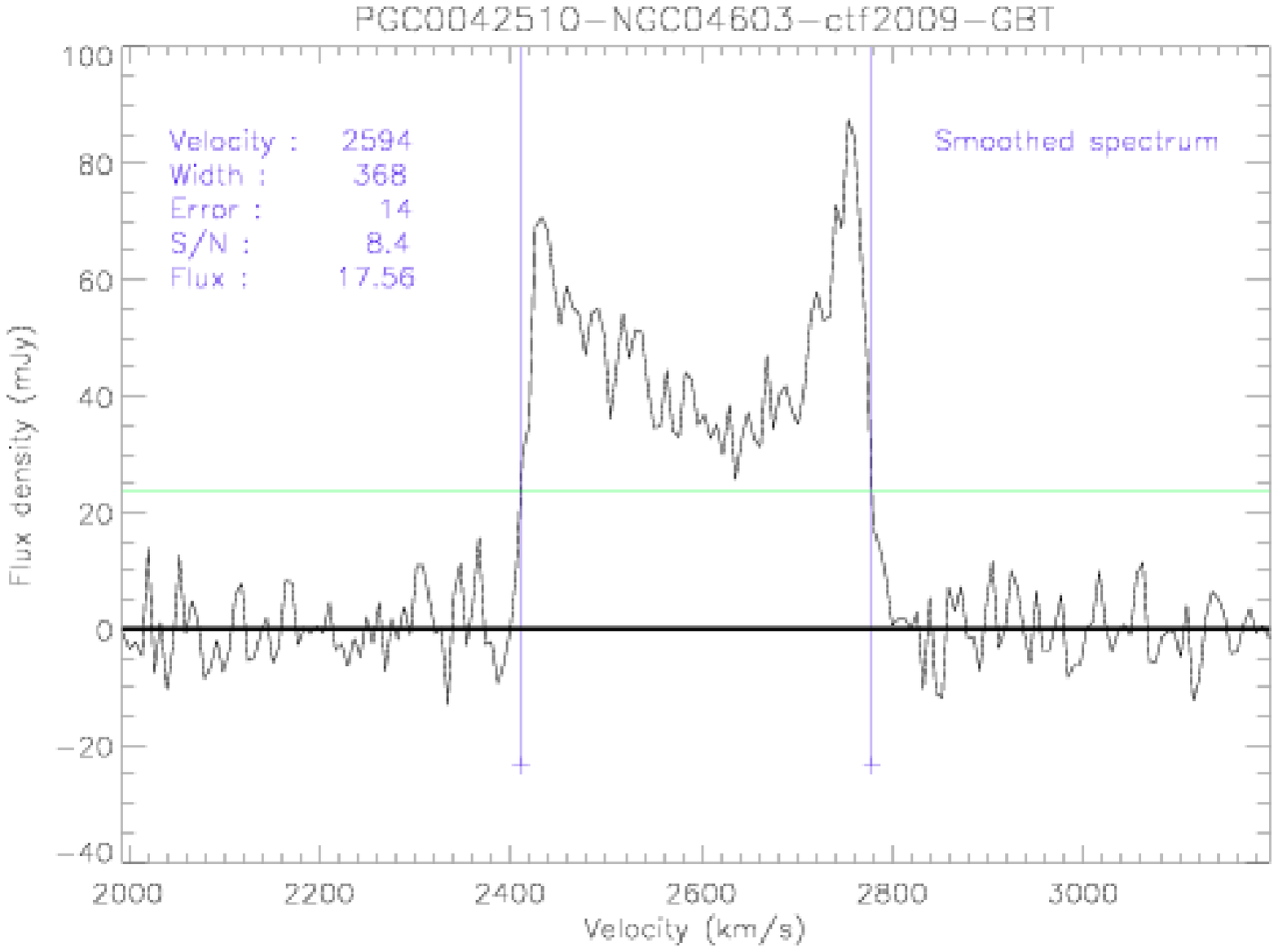}
\caption{An image of the galaxy PGC42510=NGC4603 and an HI profile of the galaxy obtained with GBT.}
\label{n4603HI}
\end{figure}

The line profile information, from whatever telescope and whatever source,  is accumulated in EDD in the catalog called {\it All Digital HI}.  The HI profiles and our fits for the parameter $W_{m50}$ are available for inspection at that site.  There can be up to three independent profiles for a given target.  The preferred measure of $W_{mx}$ is derived from an average of all contributions with $e_W \le 20$~\kms, weighted by the inverse square of the error estimates.  As we close out acquisitions for the present sample the {\it All Digital HI} catalog contains entries for 14,219 galaxies and contains entries with $e_W \le 20$~\kms\ for 11,343 galaxies.  Figure~\ref{n4603HI} is an example of the graphic material made available for every galaxy in EDD.

The {\it All Digital HI} catalog is incomplete in one important respect that requires us to retain a legacy of the analog $W_{20}$ profile measurements.  Very large galaxies or special cases such as galaxies with profiles near zero velocity confused by local emission are not adequately observed with large single dish radio telescopes.  In a small number of cases it is necessary to interpret profiles observed with interferometers or fall back to old observations with a small telescope like Dwingeloo.  In such cases, we make use of a tight correlation between the old analog parameter $W_R$ that approximates twice the maximum rotation (before de-projection to edge-on) and the new $W_{mx}$ parameter to recover a $W_{mx}$ proxy
\begin{equation}
W_{mx} = 1.015W_R - 11.25 .
\label{Wanalog}
\end{equation}

If $S/N$ is sufficient, comparable and consistent profile information is obtained with all seven radio telescopes providing data.  However some facilities are more sensitive than others.  Arecibo Telescope is the most sensitive.  As a consequence our sample extends dramatically in the declination range $0^{\circ} < \delta < 38^{\circ}$ accessed by this telescope.  GBT is the second most sensitive facility and allows observations over the entire northern sky down to $\delta = -45^{\circ}$.  Parkes Telescope is used for targets below this limit.  The lesser sensitivity of the Parkes Telescope results in reduced redshift reach toward the southern celestial pole.  Since Parkes Telescope is the smallest of current facilities it has the largest beam so observations with this telescope are favored for large nearby galaxies.

\subsection{Photometry}

The rotation curve information that has been discussed characterizes the total mass within the observed domain of galaxies.  Photometry provides a characterization of the mass in stars.  It also provides a handle on galaxy inclinations, information needed for the de-projection of
disk motions and for estimates of reddening.  After accounting for the effects of tilt, the tightness of the TFR is an evident manifestation of a correlation between stellar and total mass over the observed radii in galaxies.

Photometry may be usefully acquired over a range of wavelengths.  Most of the stellar mass is in cool stars so it can be anticipated that the best TFR correlations are in bands that are least contaminated by flux from young populations, emission lines, and dust.  With ground-based observations it has been demonstrated that the TFR scatter is minimal at $R$ and $I$ bands \citep{1988ApJ...330..579P, 2000ApJ...533..744T}.  Scatter increases at $B$, presumably because of the stochastic incidence of star formation and increased obscuration.  Scatter also increases in the near-infrared, at $H$ and $K$.  A primary cause is high and variable sky foreground which limits observations to the high surface brightness components of galaxies.  A secondary, inevitable cause of degradation results from the steepening of the TFR toward longer wavelengths \citep{1982ApJ...257..527T}.

\begin{figure}[h!]
\includegraphics[scale=0.32, angle=270]{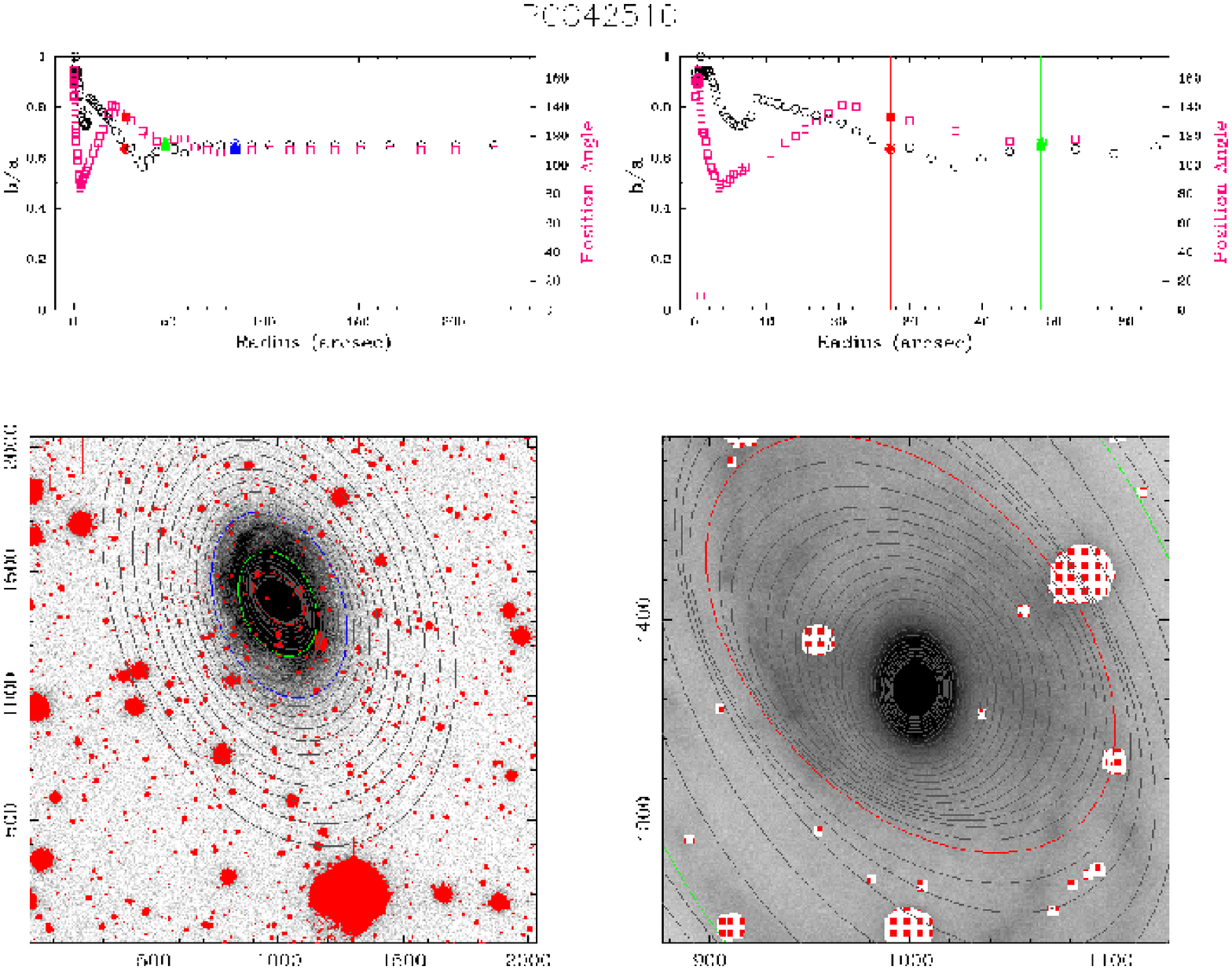}
\includegraphics[scale=0.25, angle=0]{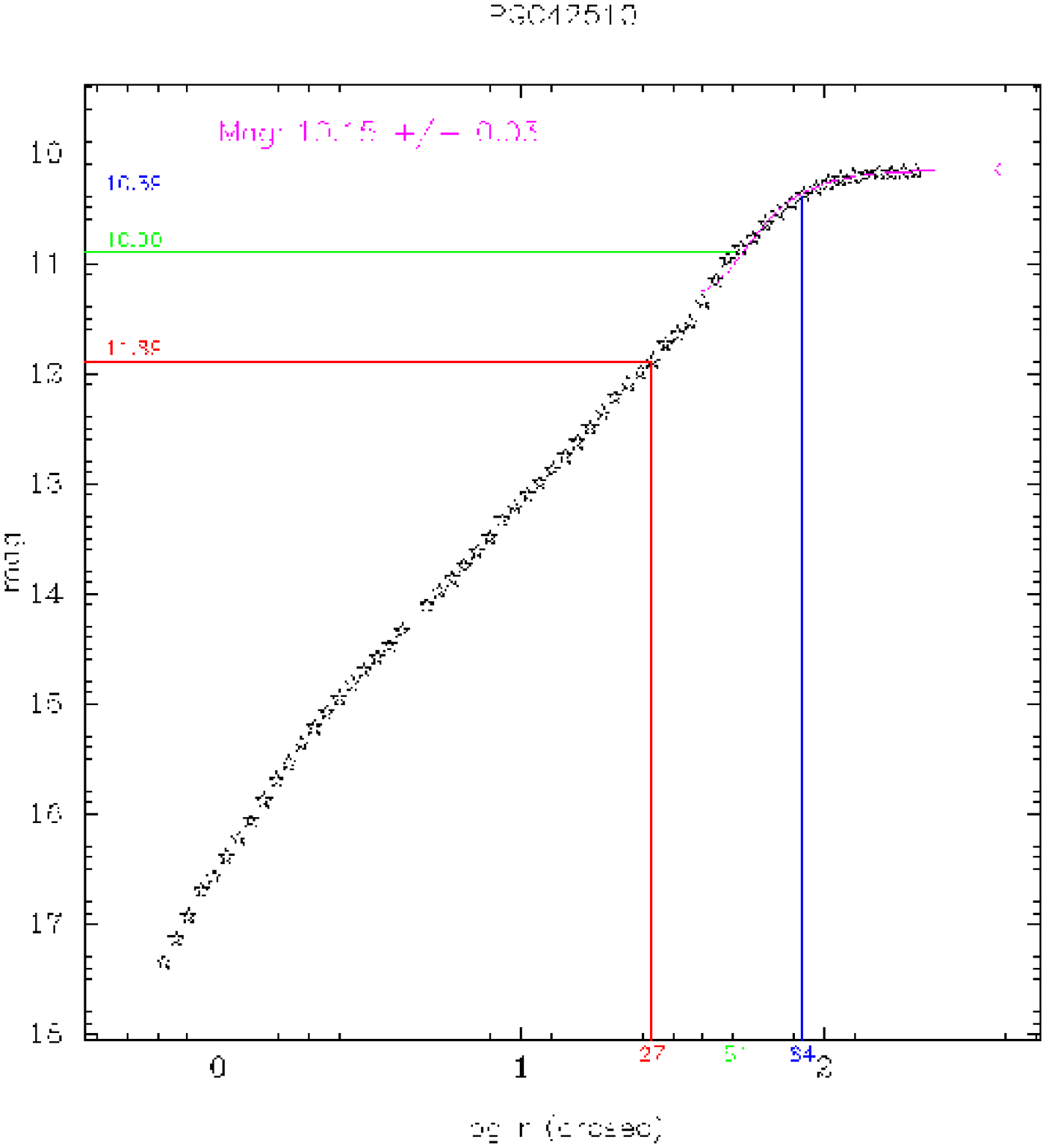}
\includegraphics[scale=0.45, angle=0]{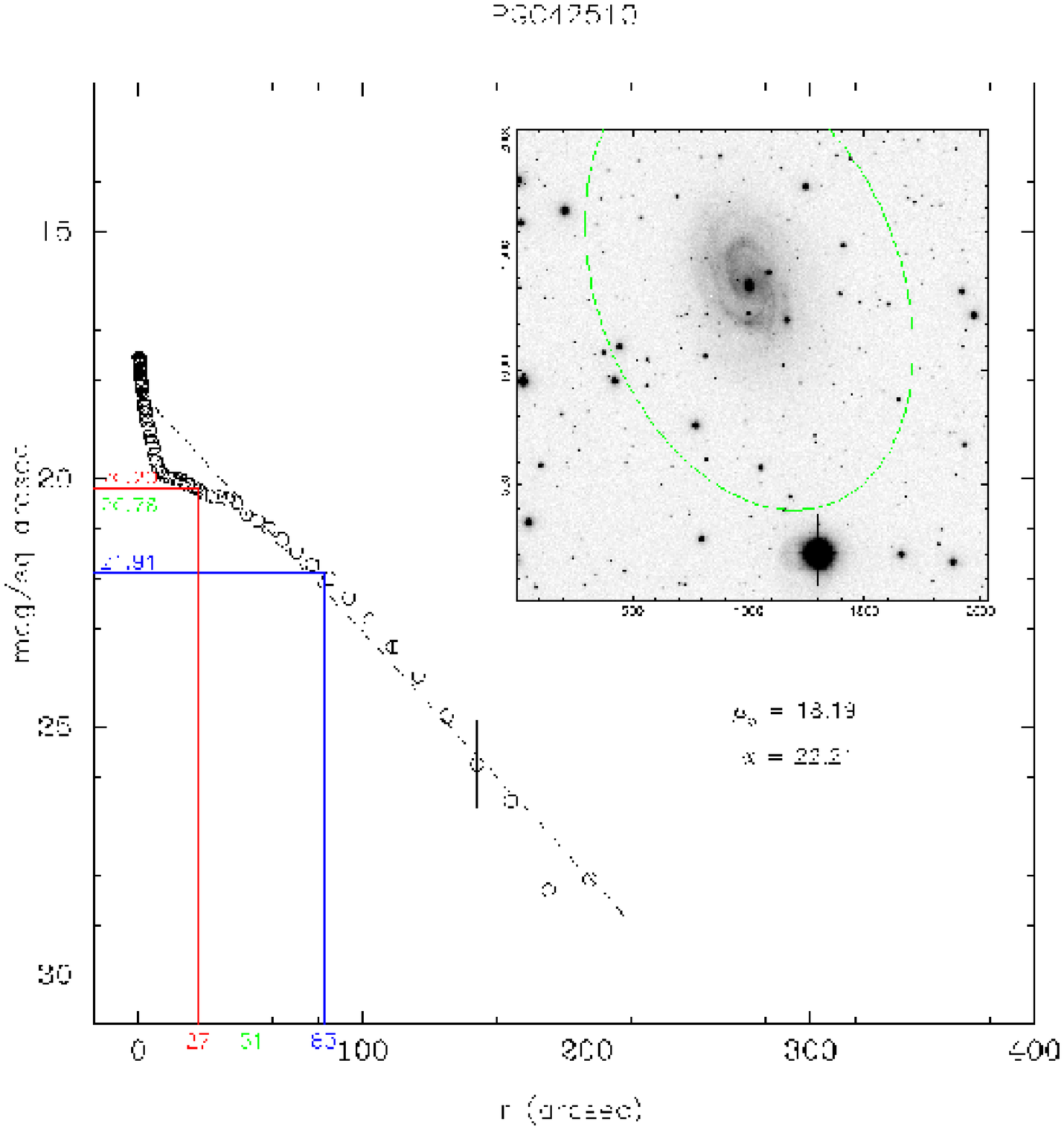}
\caption{An example of an Archangel photometry product in the EDD database.  From the top: The run of axial ratio and position angle with radius; masked images of the target on two scales; the magnitude growth curve with 20, 50, and 80\% intercepts in red, green, and blue; and the surface brightness profile with an exponential disk fit.  The galaxy in this case, NGC4603, is the same galaxy used for the HI profile illustration in Fig.~\ref{n4603HI}.}
\label{n4603ph}
\end{figure}

For the current analysis we concentrate on photometry at Cousins $I$ band, although our final product will be informed by photometry at $3.6~\mu$m obtained with Spitzer Space Telescope.  As with the TRGB and HI data sets that have already been described, we combine photometry from our own observations with published material.  Our photometry procedures were described by \citet{2011MNRAS.415.1935C} and our $I$ band data products are made available in EDD within the catalog {\it Hawaii Photometry}.  The analysis uses the Archangel photometry software \citep{2007astro.ph..3646S, 2012PASA...29..174S}.  Our products, in addition to total $I$ band magnitudes, include disk scale lengths and central surface brightnesses, radii enclosing 20, 50, and 80\% of light, a concentration index, and a ratio of dimensions on the minor and major axes.  An example of the graphic material in EDD is seen in Figure~\ref{n4603ph}.

Considerable attention was given to assure consistency with literature contributions to the $I$ band photometry compilation.  Only sources that offered substantial contributions were considered so that there would be significant overlap between samples.  We began with the assembly of magnitudes already used in {\it Cosmicflows-1} \citep{2000ApJ...533..744T, 2008ApJ...676..184T}.  The most significant contributions in that earlier compilation were from \citet{1992ApJS...81..413M}, \citet{1997AJ....113...22G}, and our own program \citep{1988ApJ...330..579P,1996AJ....112.2471T}.  Now, in addition to the new photometry described by \citet{2011MNRAS.415.1935C}, we include the accumulated and reanalyzed photometry from the Cornell group compiled by \citet{2007ApJS..172..599S} and the photometry carried out on Sloan Digital Sky Survey material by \citet{2012MNRAS.425.2741H}.  This latter contribution involves observations at Gunn $i$ band so requires a translation from Sloan $g,r,i$ to Cousins $I$ band as prescribed by \citet{2002AJ....123.2121S}
\begin{equation}
I_c^{sdss} = i - 014(g-r) - 0.35
\end{equation}
where cases with $r-i \ge 0.95$ are excluded.  After making these translations, a slight tilt was found in comparisons between $I_c^{sdss}$ and $I_c$ from the three alternative sources mentioned above ({\it Cosmicflows-1}, Cornell, or recent Hawaii).  From 725 cases in common with {\it Cosmicflows-1} and 857 cases in common with the Cornell compilation the following empirical correction was determined
\begin{equation}
I_c = 1.017 I_c^{sdss} - 0.221.
\end{equation}
The uncertainty on the slope is $\pm 0.003$ so the deviation from slope unity is significant at $5.5 \sigma$.

After these adjustments, all contributions were determined to be on the same scale.  In cases of targets with multiple measurement, all photometry sources were regarded as equal in taking an average.  On instances there were strongly deviant measures.  With 3 or more independent determinations the bad measures could be unambiguously culled.  With two determinations it was often obvious which contribution was bad.  Otherwise, if the difference was not extreme the two determinations were averaged.  The relative quality of the sources could be evaluated by pairwise comparisons.  The rms scatter in the differences between the {\it Cosmicflows-1} and Cornell sources with 2106 cases (rejecting 20 differences $>0.45$) was 0.077 mag.  The scatter between {\it Cosmicflows-1} and SDSS with 713 cases (after 12 rejections) was 0.117 mag.  The scatter between {\it Cosmicflows-1} and the Hawaii photometry with 223 cases (after 4 rejections) was 0.120 mag. 

A raw observed magnitude requires adjustments to account for obscuration in our Galaxy, $A^b_I$, and in the target galaxy, $A^i_I$, as well as a small $k$-correction that accounts for spectral displacement with redshift, $A^k_I$.  To summarize, the corrected magnitude is
\begin{equation}
I_T^{b,i,k} = I_T - A_b^I - A_i^I - A_k^I
\label{IT}
\end{equation}
where $I_T$ is the total observed magnitude and the three correction factors are 

$A_b^I = R_I E(B-V)$ with differential reddening $E(B-V)$ from \citet{1998ApJ...500..525S} cirrus maps and $R_I = 1.77$,

$A_i^I = \gamma_I {\rm log}(a/b)$ with $a/b$ the major to minor axis ratio and $\gamma_I = 0.92 + 1.63({\rm log}W^i_{mx} - 2.5)$ or $\gamma_I=0$ if $W^i_{mx} \le 86$~\kms\ \citep{1998AJ....115.2264T},

$A_k^I = 0.302z + 8.768z^2 - 68.680z^3 + 181.904z^4$ \citep{2010MNRAS.405.1409C}.

\begin{figure}[h!]
\includegraphics[scale=0.45, angle=0]{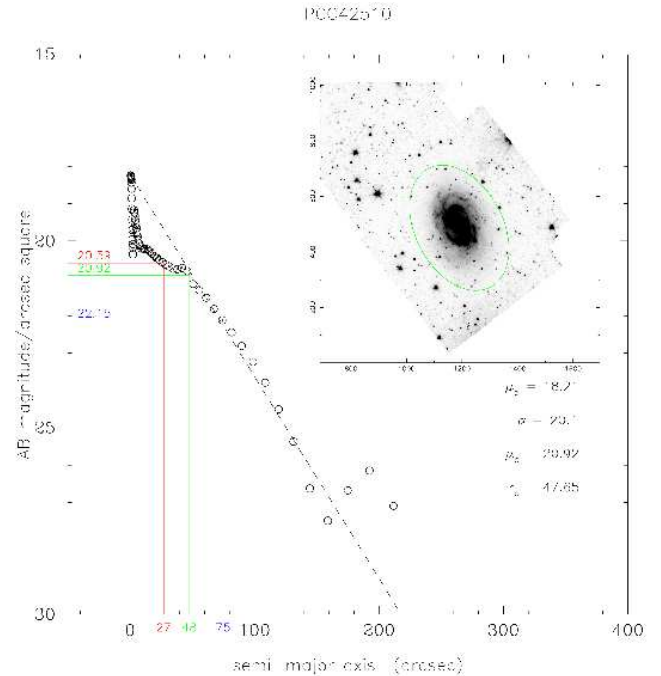}
\caption{An example of Archangel photometry of a Spitzer image.  The galaxy NGC4603 is the same as used in previous figures.  This surface brightness profile at $3.6 \mu$m can be compared with the profile at $I$ seen in Fig.~\ref{n4603ph}.}
\label{n4603L}
\end{figure}

It remains here to mention our use of Spitzer Space Telescope $3.6 \mu$m photometry.  Our procedures are described by \citet{2012AJ....144..133S}.  Again we resort to a combination of observations made within our collaboration and observations by others made available through the Spitzer archive.\footnote{http://irsa.ipac.caltech.edu/data/SPITZER/docs/spitzerdataarchives/}  In this case all observations are made with the same facility and with very similar observing strategies.  Again, reductions were carried out with the Archangel software and products are made available in EDD, this time in the catalog {\it Spitzer [3.6] Band Photometry}.  See Figure~\ref{n4603L} for an example of a surface brightness profile.

The Spitzer mid-infrared photometry has a couple of clear advantages.  It is a concern with the $I$ band photometry that the assembly is a composite of material acquired at several observatories, with subtly different filters and detectors, over different parts of the sky, many observing seasons, and with the vagaries of sky conditions.  The Spitzer photometry, by contrast, is consistently obtained all-sky.

In addition to that most important advantage, the Spitzer photometry profits from being negligibly affected by obscuration, by being sensitive to low surface brightness features because of minimal background, and because the target fluxes are dominated by light from old stars.  As a consequence of these advantages the Spitzer magnitudes {\it after corrections} have uncertainties of only 0.05 mag.  At this level magnitude uncertainties are a minor contributor to the TFR error budget.

The Spitzer photometry will lead to an important improvement of the TFR in the future.  At present the number of galaxies with reduced Spitzer photometry is limited.  The TFR calibration with currently available material will be discussed in a later sub-section and this calibration is important for the establishment of the overall distance scale.  However, by far the larger compilation of photometry today is at $I$ band.  The TFR component of {\it Cosmicflows-2} is a construct at $I$ band with only a minor zero point adjustment arising from the Spitzer calibration.

\subsection{Inclinations}

The final important ingredient needed for a TFR distance determination is a decent inclination.  Uncertainties in inclination can be a dominant source of observational error.  There are two regimes sensitive to inclination effects: those of increased obscuration toward edge-on and increased projection uncertainty in rotation velocities toward face-on.

The issue of obscuration toward edge-on is the lesser problem.  Corrections at optical bands can be quite large but they are remarkably predictable.  It is well demonstrated that obscuration is greatest in the most intrinsically luminous spirals, while at the extreme of the smallest spirals obscuration effects are marginally detected \citep{1995AJ....110.1059G, 1998AJ....115.2264T}.  The coupling of reddening with luminosity is a problem since the measurement of distances is inextricably linked with the measurement of luminosities.  \citet{1998AJ....115.2264T} sidestep the connection between reddening and luminosity by the replacement connection between reddening and linewidth, the latter a distance independent monitor of intrinsic luminosity.  The prescription for reddening at $I$ band, $A_i^I$, was already given in the qualifications for Eq.~\ref{IT}.  At the Spitzer $3.6 \mu$m band the reddening, $A_i^{[3.6]}$, is small.  The coefficient that describes the amplitude as a function of axial ratio is $\gamma_{[3.6]} = 0.10 + 0.19({\rm log}W^i_{mx} - 2.5$) or $\gamma_{[3.6]}=0$ if $W^i_{mx} \le 94$~\kms\ (Sorce et al. 2013).  Reddening in magnitudes at $3.6 \mu$m is reduced by a factor 9 from $I$ band. 

There can be greater problems with corrections to linewidths for galaxies seen toward face-on.  Some galaxies are sufficiently regular that their orientations can be determined with precisions of $1^{\circ}-2^{\circ}$.  However others present difficulties.  The tilt of barred systems can be ambiguous and likewise that of some prominent spirals, depending on the orientation of these features with respect to the major and minor axes.  Then there are galaxies with asymmetries and warps that still seem reasonable candidates for a TFR application.   Overall, comparisons between axial measurements by different authors and tests that involve sorting images by inclination suggest that a typical uncertainty in inclination is $5^{\circ}$. Axial ratios can be a poor descriptor of inclinations with large-bulge systems, witness the Sombrero galaxy. 

This level of uncertainty prescribes the limit we set of $45^{\circ}$ for our TFR samples.  An uncertainty of $5^{\circ}$ at this limit amounts to an uncertainty in linewidth of 9\%.  For all the concern that we have about inclinations, it is comforting that with tests involving cluster samples (galaxies assumed to be at the same distance) discussed by \citet{2012ApJ...749...78T} there are no TFR systematics or increase in scatter over the full range of inclinations from $45^{\circ}$ to $90^{\circ}$.  Nonetheless, the issue of inclination corrections is sufficiently disconcerting that we are attracted to a particular sample that eliminates the problem: galaxies from the various incarnations of the Flat Galaxy Catalog
\citep{1993AN....314...97K, 1999BSAO...47....5K, 2004BSAO...57....5M, 2003A&A...407..889K, 2009Ap.....52..335K}.  These galaxies are all seen extremely edge-on.  They constitute a particularly interesting sample when combined with Spitzer photometry where reddening is minor.  We consider such a sample, as will be discussed.

We define inclinations, $i$, from measurements of axial ratios, $b/a$, the minor to major axis dimensions, with the formula
\begin{equation}
{\rm cos}~i = \sqrt{{(b/a)^2 - q_0^2}\over{1 -q_0^2}}
\label{q0}
\end{equation}
where $q_0=0.2$ is the assumed flattening of an edge-on system.  Other authors use involved variants of $q_0$, noting that some edge-on systems are distinctly thinner than $b/a=0.2 $\citep{1983A&A...118....4B, 1997AJ....113...22G}.  However we prefer this simpler formulation, recalling that our primary interest is not the inclination per se but rather the correction needed to measure a distance.  This simple formulation avoids discontinuities.  Also, as discussed at length by \citet{2000ApJ...533..744T}, a variation in $q_0$ translates to an almost fixed displacement of linewidth at all inclinations: an uncertainty in the numerator on the right side of Eq.~\ref{q0} because of the choice of $q_0$ has a large effect on $i$ toward edge-on but a small effect on the linewidth correction $1/{\rm sin}~i$, whereas the uncertainty in this numerator because of choice of $q_0$ is small toward face-on where the potential effect on $1/{\rm sin}~i$ would be great.

Our sources of axial ratios tracks our sources of $I$ band photometry with one important extension. We find that the axial ratios given in the Lyon Extragalactic Database (LEDA) are of good quality  \citep{1996A&A...311...12P}.   Tiny adjustments were made to arrive at consistency with $b/a$ values used earlier in our program:

$b/a = b/a^{cornell} - 0.01$,

$b/a = 0.95 b/a^{sdss} + 0.01$,

$b/a = 0.97 b/a^{leda} + 0.01$.

Inter-comparisons with the large Cornell sample  \citep{2007ApJS..172..599S} result in the following rms scatter in differences: $\pm 0.04$ (582 cases) with {\it Cosmicflows-1}, $\pm 0.05$ (810 cases) with SDSS, $\pm 0.07$ (167 cases) with the Hawaii photometry, and $\pm 0.08$ (2104 cases) with LEDA.  The LEDA contribution assures that there are at least two independent $b/a$ measurements for all galaxies with $I$ band photometry.  The availability of multiple measures is helpful for the culling of bad data.  When the occasion demanded it, we introduced new measurements of $b/a$ from our evaluation of images.

\subsection{The TFR Calibration} 

The TFR calibration used with {\it Cosmicflows-1} dates from \citet{2000ApJ...533..744T}.  An update is required, especially because of the new definition of the HI linewidth parameter.  The required re-calibration for the TFR at $I$ band was published by \citet{2012ApJ...749...78T}.   A parallel calibration for the TFR using Spitzer [3.6] band photometry is given by \citet{2013arXiv1301.4833S}.   The procedures followed for the construction of the TFR are described in detail in these three references so only a brief outline of important points will be given here.

In order to minimize the Malmquist `selection' bias \citep{1994ApJS...92....1W} we derive distances with the `inverse' TFR, the linear fit to the correlation between magnitudes and log linewidth with errors taken in linewidth.  Only a tiny correction is required for a residual bias, as will be discussed later.  A feature of the inverse relation (and a check that it is minimally biased) is invariance of the slope with the limiting magnitude of the sample.  Hence a template can be constructed through the superposition of subsamples drawn from clusters, with appropriate shifts on the magnitude axis to account for relative differences in distance moduli.

\begin{figure}[h!]
\includegraphics[scale=0.45, angle=0]{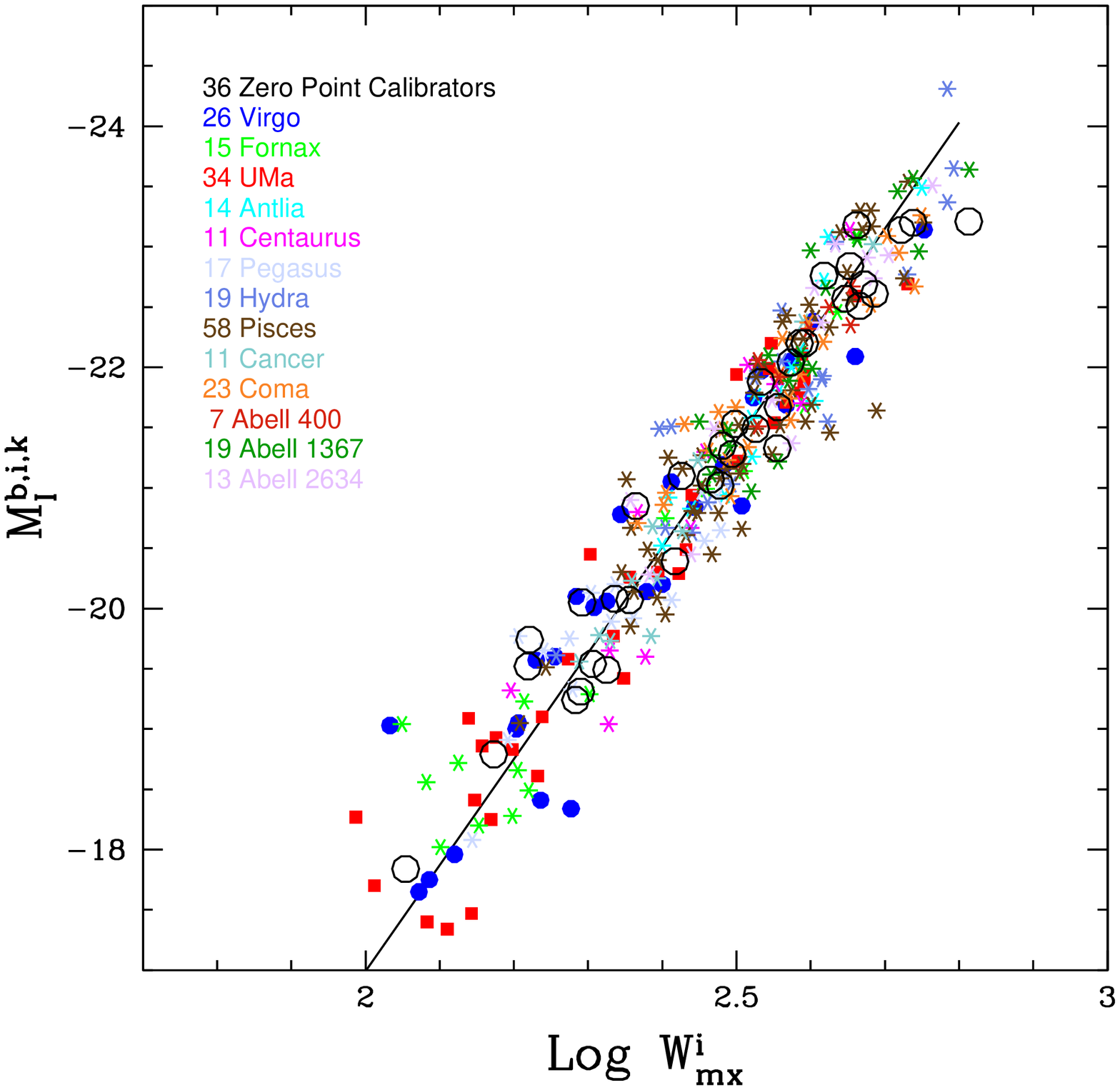}
\includegraphics[scale=0.45, angle=0]{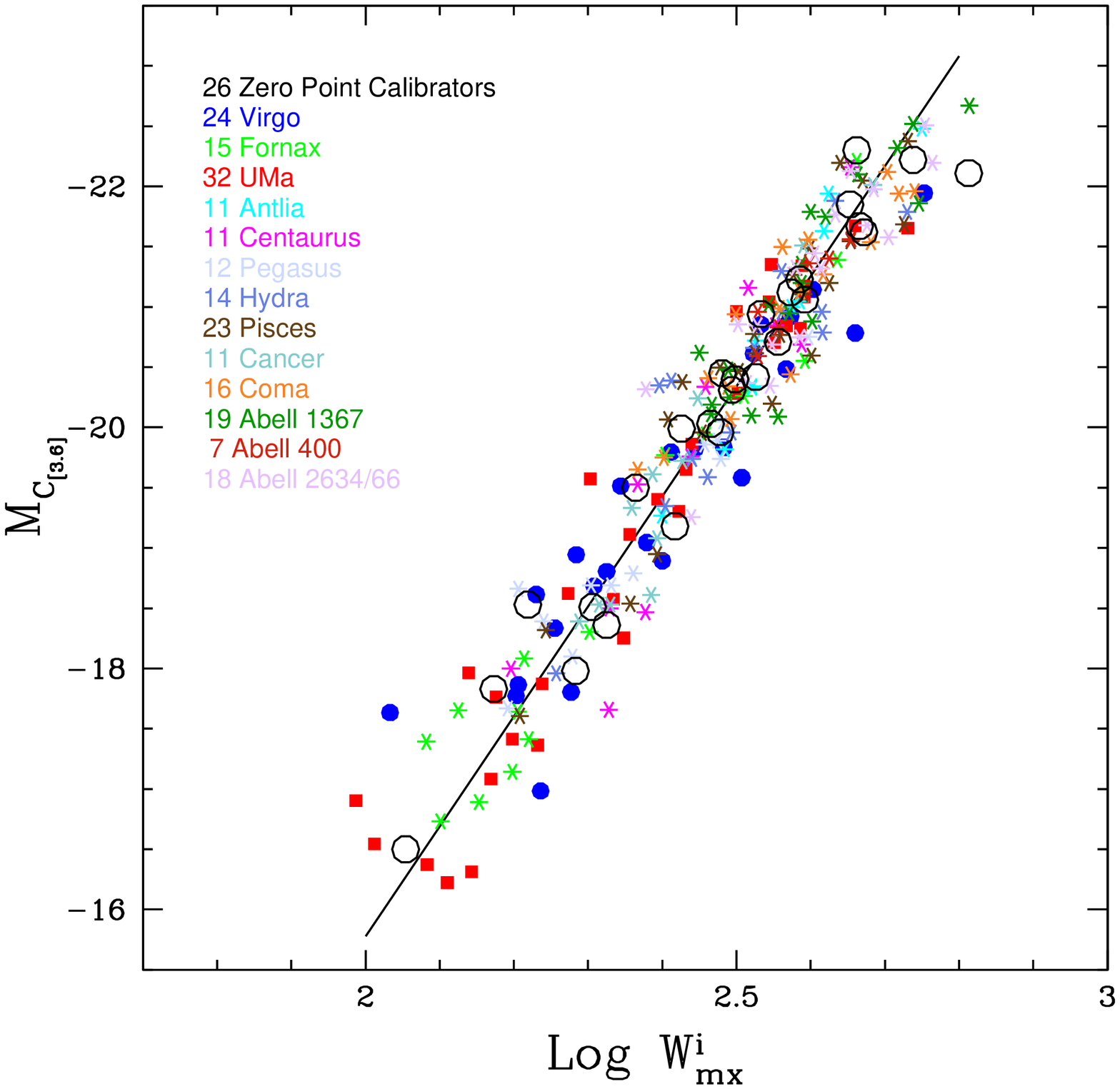}
\caption{Two calibrations of the correlation between galaxy luminosity and HI linewidth. {\it Top:} $I$ band calibration (Vega magnitudes). {\it Bottom:} Spitzer [3.6] calibration (AB magnitudes).  Colored symbols: galaxies in 13 different clusters distinguished by different colors and symbol types, slide in magnitude to a best fit.  These `template' galaxies define the slopes of the solid lines.  Large open circles: galaxies with Cepheid PLR or TRGB distances that set the zero points of the solid lines.}
\label{TFR}
\end{figure}

With the current investigation we draw on observations within 13 clusters.  A rapidly converging iteration between choice of slope and modulus shifts between the 13 clusters leads to an optimal matching of the clusters and a definition of the inverse TFR slope.  The final step is to define the zero point using galaxies with distances independently established by Cepheid PLR or TRGB observations.  The correlation slope determined with the 13 cluster template is assumed in deriving the least squares best fit with the zero point calibrators.  The results of these procedures are demonstrated for the $I$ band calibration by \citet{2012ApJ...749...78T} in the top panel of Figure~\ref{TFR}.

There is a complication with the calibration at $3.6 \mu$m.  There is a color term.  The rms magnitude dispersion of the $I$ and [3.6] correlations are comparable at $\sim0.4$ mag but only after correction for a color term in the mid-infrared case.  The color dependency arises because of the well known correlation between galaxy morphology and color that causes  the TFR to steepen in bands toward longer wavelengths \citep{1982ApJ...257..527T}.  Given two galaxies of different colors but the same linewidth, the two galaxies must displace in magnitude in different passbands.  Empirically, displacements are minimum around $1 \mu$m.  At shorter wavelengths the bluer of our hypothetical pair will tend to be brighter while at longer wavelengths the redder system becomes brighter.  $I$ band measurements are near the inflection wavelength so there is no clear advantage for the introduction of a third parameter adjustment but by the mid-infrared there is a clear need for a color adjustment.  The issue of a third parameter has been debated in the literature.  It is generally couched as a surface brightness or morphological dependence \citep{1985ApJ...289...81R, 1998A&AS..130..333T, 2006ApJ...653..861M} but surface brightness and morphology are closely correlated with color.  The use of morphology rather than color as an additional parameter creates unfortunate discontinuities in the distance tool.  Also, morphology assignments are subjective and may vary with distance.  An alternative, more quantitative stand-in for morphology makes use of an HI flux to near infrared luminosity ratio  \citep{2008Ap.....51..336K}; later types have relatively higher HI flux ratios.

The [3.6] band TFR with the color correction that was derived by \citet{2013arXiv1301.4833S} is shown in the lower panel of Figure~\ref{TFR}.  The slope template is based on the same 13 clusters, using the same galaxies in all cases where Spitzer photometry is available (213 of the 267 galaxies in the $I$ calibration).  The zero point is set by the subset of galaxies with Cepheid PLR or TRGB distances with Spitzer photometry (26 of the 36 galaxies in the $I$ calibration).  The calibrations in the two bands use the same linewidth and inclination values and the same reddening procedures save for the transparency differences with wavelength.

The results of the TFR calibrations are summarized with the following relations:
\begin{equation}
M_I^{b,i,k} = -21.39 - 8.81 ({\rm log} W^i_{mx} - 2.5)
\end{equation}
with $1\sigma$ uncertainties of $\pm0.07$ in the zero point and $\pm0.16$ in the slope and magnitudes in the Vega system.
\begin{equation}
M_{C_{[3.6]}} = -20.34 - 9.13 ({\rm log} W^i_{mx} - 2.5)
\end{equation}
with $1\sigma$ uncertainties of $\pm0.08$ in the zero point and $\pm0.22$ in the slope.  The pseudo-magnitude $M_{C_{[3.6]}}$ is an absolute magnitude in the [3.6] band where the apparent magnitude after color correction is
\begin{equation}
C_{[3.6]} = [3.6]^{b,i,k,a} + 0.47 [(I - [3.6]) + 0.77].
\end{equation} 
The [3.6] magnitudes are in the AB photometric system.

There is a small Malmquist selection bias to distances that requires correction in either band.  It arises for two reasons.  First, the magnitude cutoffs in the cluster samples were made at constant values in the blue but the cutoff then slants with linewidth at redward wavelengths since larger linewidth galaxies are redder.  Second, as a consequence of the exponential cutoff in the galaxy luminosity function more galaxies are available to scatter brightward than scatter faintward, especially at large distances where most candidates are near the bright limit.  A correction $b$ to distance modulus $\mu$ is required
\begin{equation}
b = \beta_{\lambda} (\mu -31)^2
\end{equation}
where $\beta_I = -0.005$ and $\beta_{[3.6]} = -0.0065$.  

With either version of the TFR calibration offered here, the magnitude rms scatter is 0.4 so a distance can be measured to a target with an uncertainty of 20\%.  The scatter is reasonably approximated by a gaussian in the core but there are non-gaussian wings, with $\sim 2\%$ of candidates manifesting scatter greater than $3\sigma$.  These are cases that escaped screening during the processing of sample selection, profiles, photometry, and inclinations so the reasons for deviations are often not clear.

\subsection{TFR Samples}

The considerable contribution of archival material is in itself an inchoate sample.  With our own observations as a supplement we have focused on five interconnected programs.  One of these has already been discussed: the assembly of the data required for a construction of the TFR cluster template and zero point calibrators.  A second gave attention to spiral galaxies that have hosted SNIa in order to reinforce the bridge between TFR and SNIa distances.  Current results with this program have been published by \citet{2012ApJ...749..174C} and \citet{2012ApJ...758L..12S} and will be given further attention in the next section.

The other three programs involve large all-sky samples designed to improve our knowledge of velocity fields in successive outward steps.  The three programs were discussed in some detail by \citet{2011MNRAS.414.2005C} so we only need to summarize here.

The first of the three aims to achieve close to completion in coverage of applicable large spirals within 3000 \kms.  The word `applicable' is in recognition that there are systems too face-on, or confused, or disrupted to be considered for TFR analysis.  The limit of 3000~\kms\ is somewhat a legacy of very early work \citep{1981ApJS...47..139F} conditioned by the sensitivity of radio telescopes of the day.  The issue of sensitivity is still germane.   A spiral galaxy within 3000~\kms\ is now easily detected at high $S/N$ wherever it lies in the sky.   The 3000~\kms\ limit nicely includes the entire historic Local Supercluster \citep{1953AJ.....58...30D} and its limit just reaches the next important structures including the Centaurus and Hydra clusters (straddling the 3000~\kms\ boundary but fully included as components of the cluster template).  Specifically, we seek to achieve completion with all galaxies within 3300~\kms\ with $M_{K_s} < -21$, $i>45^{\circ}$, type later than Sa, and not confused or distorted.  The absolute magnitude cut comes from the observed 2MASS magnitude \citep{2003AJ....125..525J} and redshift interpreted through a preliminary model of local flows.  This program builds on {\it Cosmicflows-1} that already gave high density coverage of the Local Supercluster.  {\it Cosmicflows-2} now contains TFR distance estimates for 1360 galaxies with $V_h < 3300$~\kms.

The second extensive sample was built with two principal considerations.  One is a recognition of the major structures in the Centaurus-Hydra-Norma \citep{1987ApJ...313L..37D} and Perseus-Pisces \citep{1988lsmu.book...31H} regions that lurk just beyond the 3000 \kms\ limit of the nearby sample.  We wanted these important structures to be included in the extended sample.  The other consideration is the sensitivity of radio telescopes today.    If a target enters our preliminary sample we want to expect with high probability that it will be detected with acceptable $S/N$ in HI.  For these reasons, our second sample is limited to $V_{cmb} < 6000$~\kms.  The candidates are selected from the PSCz redshift survey of galaxies brighter than 0.6 Jy at $60 \mu$m detected with IRAS, the Infrared Astronomical Satellite \citep{2000MNRAS.317...55S}.  There is an inclination limit, $i>45^{\circ}$, and a color limit, $60-100 \mu$m flux less than unity, the latter to discriminate against starburst systems in favor of normal spirals with emission predominantly from cirrus.  With the selection based on far-infrared flux the sample is minimally affected by obscuration (except in the ability to identify sources at extreme low latitudes) so we get good coverage toward the Galactic plane.  The Centaurus-Hydra-Norma and Perseus-Pisces regions both flirt with the zone of avoidance.  The galaxies that have been observed randomly sample within the selection criteria that have been discussed.  Presently we have distance estimates for 1363 galaxies with $3300<V_h<6000$ \kms.  We have less than satisfactory coverage at $\delta < -45^{\circ}$, the exclusive domain of Parkes Telescope.  A continuing effort is being made to improve this situation. 
 
A third sample extends to greater redshifts in order to investigate the nature of galaxy flows on very large scales.  It draws from the Revised Flat Galaxy Catalog \citep{1999BSAO...47....5K}.  The thinnest galactic systems are spirals of type Sc with small bulges and well constrained disks.   These properties give targets selected from the Flat Galaxies Catalog a homogeneity and amenability to HI detection that is favorable for a sparse all-sky sampling \citep{2009Ap.....52..335K}.  At present the implementation is instrumentally restricted.  Most of the currently observed galaxies in the flat galaxies program with large velocities lie in the declination range of the Arecibo Telescope and have been observed with that facility.  GBT has contributed over a wide range of declinations but generally with targets at lower redshifts.  {\it Cosmicflows-2} includes 1446 galaxies at $V_h > 6000$~\kms\ largely drawn from the Flat Galaxy Catalog.  Figure~\ref{histVtf} gave a histogram that summarizes the redshift distribution of our entire TFR contribution to {\it Cosmicflows-2}.  The peaks within 3000~\kms\ and 6000~\kms\ are reflections of prominent over-densities in the Local Supercluster, Centaurus-Hydra-Norma, and Perseus-Pisces convolved with the emphasis of our samples.  \citet{2011MNRAS.414.2005C} provide a more detailed breakdown of the sample coverage.

\section{Integration of Methodologies}

We are dealing with six primary methods of measuring distances: the Cepheid PLR, TRGB, SBF, TFR, FP, and SNIa.  Within each method it is necessary to reconcile the contributions from different practitioners.  As best we can, we need to assure that these large numbers of components are measuring distances in a consistent manner.  There are issues of zero point and issues of linearity of scale with distance.  Fortunately there is a lot of overlap.  Our primary goal of determining peculiar velocities can be reached with only relative consistency of the zero point but, to the degree possible, we seek the correct absolute zero point.

The integration proceeds in steps moving outward in distance.  As a starting point we accept the HST distance scale key project results with the Cepheid PLR \citep{2001ApJ...553...47F} and subsequent Cepheid PLR observations with consistent methodology \citep{2011ApJ...730..119R}.  The key project zero point assumes the Large Magellanic Cloud distance modulus of 18.50.  For the moment we construct an edifice that seeks consistency within this zero point assumption.  As a final step, after a review of the latest evidence, we evaluate whether the zero point should be tweaked.

Step one was already discussed: the comparison that assures the Cepheid PLR and TRGB measurements are compatible.  The good agreement, within 0.01 mag, was shown by \citet{2007ApJ...661..815R} and by several authors in the special case of the Maser galaxy NGC4258  \citep{2007ApJ...661..815R, 2008ApJ...689..721M, 2011ApJ...730..119R}.  It is pleasing that this agreement was not forced.  The routes to the Population I Cepheid PLR and the Population II TRGB calibrations are independent, save that they ultimately are founded on trigonometric parallaxes.
Particularly the Cepheid PLR, but also the TRGB, form the pedestals for the other procedures.  

As an aside, distance determinations to galaxies within the Local Group represent special cases.  Distances to immediate satellites of the Galaxy usually involve Horizontal Branch or RR Lyr observations \citep{2009AJ....138..459P}.  Distances for satellites of M31 mostly come from ground-based TRGB work  \citep{2013ApJ...766..120C}.  Some eclipsing binary measures are becoming available \citep{2006ApJ...652..313B}.  On large scales, Maser distances are starting to become available and, in addition to the famous NGC4258 case, we include two new distances by this geometric method \citep{2013ApJ...767..155K, 2013ApJ...767..154R}.

\subsection{TFR: The Union of CF2 and SFI++}

The construction of our version of the TFR and the foundation of 36 Cepheid PLR and TRGB calibrators was reviewed in Section 3 and discussed earlier in more detail \citep{2012ApJ...749...78T, 2013arXiv1301.4833S}.  This material is given the shorthand name CF2.  Before moving on to other methods we acknowledge  and integrate a second important compendium of TFR distances, the SFI++ compilation\citep{2007ApJS..172..599S}.  It is to be appreciated that there is a substantial overlap between SFI++ and the current work in the frequent use of the same raw photometric and HI data.  However there are sufficient differences in analysis procedures to make for interesting comparisons.  We initially accept SFI++ distances as given by Springob et al.  and now check for agreement in zero point and look for any systematics.  As an aside, we note that an absolute zero point for the TFR distances by the Cornell group was given by \citet{2006ApJ...653..861M} based on the HST key project scale but with a reduced sample of Cepheid PLR calibrators.  Here we accept the SFI++ distances as relative equivalent velocities and ultimately rescale to achieve agreement with other measures.

The SFI++ collection offers distances that are directly measured or distances adjusted for bias.  There are several reasons for bias.  There is the `selection' Malmquist bias that arises if, among galaxies with the same linewidth at the same distance, the brighter are given attention but not the fainter.  Then there are the `homogeneous' and `inhomogeneous' Malmquist biases that depend on the distribution of galaxies. In the homogeneous case, for galaxies with a given measured distance, there are more cases that arrived there from scatter inward from larger true distances than arrived from scatter outward from smaller true distances.  In the inhomogeneous case, the scatter is away from high densities toward low densities.  Each of these biases produces kinematic artifacts.  \citet{2007ApJS..172..599S} make an adjustment that globally compensates for all of these factors.

We follow a different strategy.  In the case of the selection Malmquist effect, the danger is the systematic miss-measurement of distances. In the case of the distribution Malmquist effect, whether homogeneous or inhomogeneous, distance measures can be individually unbiased but care must be taken to avoid bias in the velocity field.   Our strategy is to deal with the two distinct Malmquist biases separately.   We feel that the issues of distance measurements and velocity field inferences should be kept apart.  Our goal is to measure individual distances that, while they suffer errors, are unbiased.  Therefore we do not choose to give distances that are `wrong' in a way that compensates for suspected kinematic effects.  Specifically in dealing with the distribution Malmquist problem, we believe that a proper recipe is to evaluate distances at observed redshifts rather than inferred peculiar velocities at measured distances.  Consequently we make no adjustments for the distribution Malmquist effects in our reported distances.  Then regarding the selection Malmquist problem we note that it is possible to null the bias, and if so desired even reverse the sign of it, through the choice of the slope of the TFR.  Our procedures to define optimal slopes at our $I$ and [3.6] passbands were discussed in Section 3.4, including a description of small corrections to a residual bias that we find necessary.  

\begin{figure}[h!]
\includegraphics[scale=0.44, angle=0]{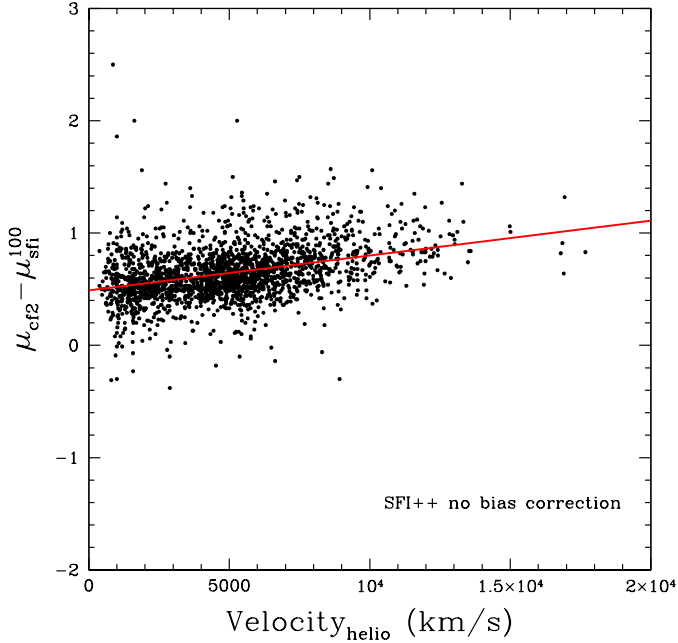}
\caption{Differences in distance moduli from our analysis (CF2) and from the SFI++ analysis for 2071 galaxies in common.  There is an offset from zero because the SFI++ distances were analyzed as nominally consistent with H$_0=100$.  The slope as a function of velocity is taken as evidence for the selection Malmquist bias in the SFI++ raw distances.}
\label{deldm}
\end{figure}

Returning to SFI++, we remark that what interests us are their directly measured distances, free of their complex corrections.  However, since a bi-variate fit that considers errors in both magnitudes and linewidths was used in the determination of their TFR slope, it is to be anticipated that the distances suffer from the selection bias.  This expectation is acknowledged by the SFI++ authors and evidence for it is found in the comparison of distance moduli for galaxies in common to our samples shown in Figure~\ref{deldm}.  After rejection of only 5 discordant differences (1 CF2 value judged to be bad and 4 SFI++) a least squares linear fit to the differences in 2071 moduli, $\Delta\mu = \mu_{cf2} - \mu_{sfi}^{100}$, has the form
\begin{equation}
\Delta\mu = 0.492(\pm0.011) + 0.000031(\pm0.000002) V_{LS}
\label{delmu}
\end{equation}
where $V_{LS}$ is the velocity of a galaxy in the Local Sheet frame \citep{2008ApJ...676..184T}, $\mu_{cf2}$ is the CF2 distance modulus with the zero point established by \citet{2012ApJ...749...78T} and $\mu_{sfi}$ is the \citet{2007ApJS..172..599S}  unadjusted modulus with a nominal zero point consistent with H$_0=100$.  After a correction of the form prescribed by Eq.~\ref{delmu} is applied to the SFI++ sample the rms agreement between CF2 and SFI++ distance moduli is $\pm0.22$ mag.  This scatter is roughly half the scatter found for cluster samples.  The agreement affirms that our distinct analysis methodologies produce comparable results and TFR scatter is dominated by factors unrelated to our procedures.

\begin{figure}[h!]
\includegraphics[scale=0.98, angle=0]{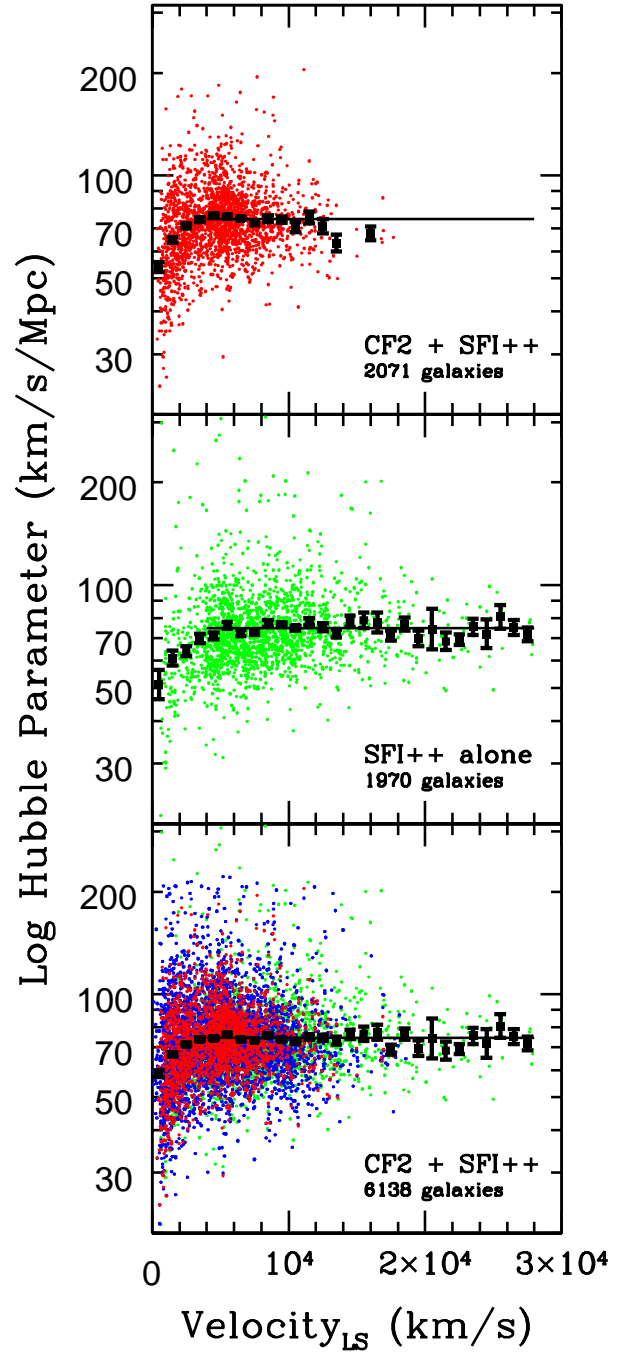}
\caption{Hubble parameter, velocity/distance, for the entire TFR sample after adjustments. {\it Top panel:} Individual points in red are based on averaged distances when both SFI++ and our new measures (CF2) are available.  Black points and error bars result from averaging in 1000~\kms\ bins.  The straight line at 74.6~\kmsMpc\ is a best fit to points with $V_{LS} > 4000$~\kms.  {\it Middle panel:} Green points represent galaxies with only SFI++ distances.  A zero point was selected to optimize the fit to 74.6~\kmsMpc\ for $V_{LS} > 4000$~\kms.  {\it Bottom panel:} The entire sample of TFR distances.  In addition to the representatives in the upper panels there are blue points for galaxies with only CF2 measures.}
\label{VH3}
\end{figure}

For each galaxy it is possible to construct its `Hubble parameter', $V_{LS}/d$, from its observed velocity and distance.  Values for this parameter are shown as a function of velocity for the galaxies with both CF2 and SFI++ distances in the top panel of Figure~\ref{VH3}.  The distances are averages of CF2 and SFI++ .  Our own CF2 distances are given double weight since we have a clearer understanding of how they were obtained, including the issue of bias treatment.  The horizontal line at H$_0 = 74.6$~\kmsMpc\ results from a fit to Hubble parameters with $V_{LS} >4000$~\kms, beyond the domain of obvious velocity perturbations.  The zero point scaling comes from \citet{2012ApJ...749...78T}.  This value of H$_0$ is not our final value; this issue will be reviewed in a later section.  Here, attention is drawn to the constancy with a specific value of the Hubble parameter with velocity.  The normal indication of a selection Malmquist bias is an increase of the Hubble parameter with redshift \citep{1993A&A...280..443T, 1994ApJ...430...13S}.  There is no hint of a residual selection bias in this data.

SFI++ includes almost 2000 galaxies not found in CF2.  It could have been supposed that the selection Malmquist bias correction implied for SFI++ from the roughly 2000 galaxies in common would serve for the SFI++ only sample but this expectation is not met.  If that correction is made then there is a highly significant {\it decrease} in the Hubble parameter with increasing velocity with the SFI++ only sample.  In the middle panel of Figure~\ref{VH3} we show the run of Hubble parameter with velocity for the SFI++ only sample raw distances with {\it no} bias correction.  The zero point has been shifted to match the best fit H$_0 = 74.6$~\kmsMpc\ for $V_{LS} > 4000$~\kms\ found in the top panel.  It is seen that with no bias  correction the run of Hubble parameter with velocity is nicely constant in the mean.

The apparent explanation for this situation derives from the reason there are this large number of systems in SFI++ that are not in CF2.  While the CF2 sample falls off abruptly at $V_{LS} \sim 12,000$~\kms, the SFI++ sample extends with significant coverage to almost 30,000~\kms.  Mostly, the difference is the inclusion of the cluster samples of \citet{1999AJ....118.1468D, 1999AJ....118.1489D}.  With these samples, rotation information was derived from optical spectroscopy rather than HI linewidths.  The merits of the optical material were discussed by \citet{2005AJ....130.1037C, 2007AJ....134..334C}.  Since we have not made any attempt to integrate optical and radio spectroscopic observations, the optical kinematic based component of the global TFR sample rests solely with SFI++.  It is not clear to us why this component of SFI++ does not manifest the selection Malmquist bias but it is probably related to the fact that these galaxies were analyzed in the context of cluster memberships.  The identifications with cluster membership will be revisited in a later secion.

In the bottom panel of Figure~\ref{VH3} we see the entire CF2 plus SFI++ TFR sample.  In addition to the 2071 cases in common to the two sources in red and the 1970 cases added by SFI++ alone in green, there are 2097 sources provided by CF2 alone in blue for a total of 6138 galaxies.  All sub-samples have been brought to the same zero point scale resulting in a mean Hubble parameter value of 74.6~\kmsMpc\ with an rms scatter of 24.2~\kmsMpc.  After culling discussed in Section 5, the combined TFR sample retains 5998 galaxies.

\subsection{Integration of Surface Brightness Fluctuation and Fundamental Plane Distances}

The important Surface Brightness Fluctuation (SBF) sample of \citet{2001ApJ...546..681T} was already incorporated within {\it Cosmicflows-1}.  This material is again included with only the small modifications advocated by \citet{2010ApJ...724..657B}.  What is new with the SBF technique is the contribution made by HST observations which carry the promise of extending the utility of the method from $\sim40$~Mpc from the ground to $\sim100$~Mpc from space \citep{2012Ap&SS.341..179B}.  Currently, the contributions from HST are either of a calibration nature or restricted to the nearest two large clusters, Virgo and Fornax  \citep{2007ApJ...655..144M, 2009ApJ...694..556B}.  These studies provide a valuable constraint on the {\it relative} distances of these two key clusters to our dynamics.  The Virgo study clarifies the status of the Virgo W$^{\prime}$ Group in the cluster line-of-sight but 50\% more distant.  The detailed information will be of great value in subsequent dynamical modeling.

SBF studies are limited to systems with predominantly old populations and this constraint is shared with the Fundamental Plane (FP) technique.  The elliptical and S0 galaxies that are useful for FP studies are prominently represented in clusters.  The accuracy of the FP procedure is greatly enhanced by averaging over cluster members.  Consequently for the present compilation we consider contributions from three literature sources that emphasize observations within clusters.  These studies go by the acronyms SMAC, EFAR, and ENEARc \citep{2001MNRAS.327..265H, 2001MNRAS.321..277C, 2002AJ....123.2990B}.  There were already discussions of these samples, briefly in the TFR calibration paper by \citet{2012ApJ...749...78T} and more extensively in the SNIa calibration paper by \citet{2012ApJ...749..174C}.  Here we need only review.

Our goal is to assure a consistency between these three FP sources and then a consistency with the other methodologies at our disposal.  The SMAC sample serves as a good intermediary between the FP samples, with good overlap with both ENEARc and EFAR, and as an intermediary with an absolute calibration because of a literature linkage with SBF \citep{2001MNRAS.327.1004B}.  This latter connection to an absolute scale is used to a first approximation but small modifications are required to match the CF2 system.  Regarding the sample overlaps, an inter-comparison of the three FP contributions is summarized in Figure 3 of \citet{2012ApJ...749..174C}.  SMAC and ENEARc consider 19 clusters in common and SMAC and EFAR have 11 in common, with a total of 28 clusters observed by at least two teams.  Pairwise modulus correlations in each case are consistent with slope unity with distance.  Standard deviations of the fits are 0.05 mag with ENEAR$-$SMAC and 0.09 mag with EFAR$-$SMAC.  An additional 105 clusters are observed by one team only.  FP distances are available from these sources for 1508 separate galaxies.  

\begin{figure}[h!]
\includegraphics[scale=0.43, angle=0]{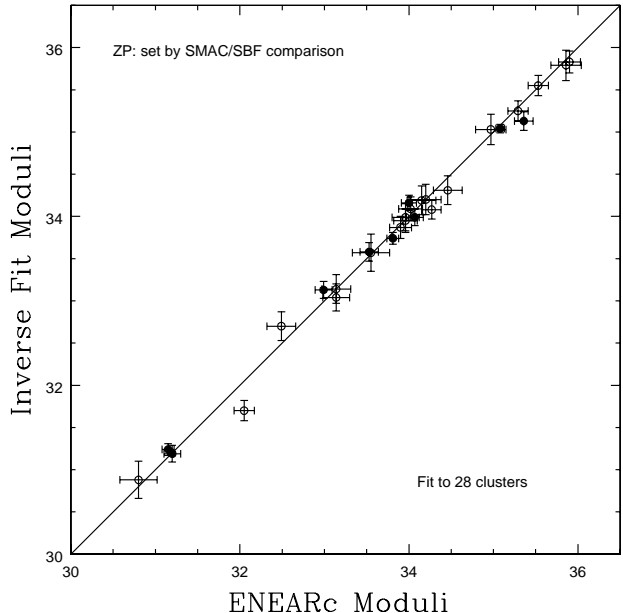}
\caption{Comparison of ENEAR literature moduli and inverse fit moduli determined with the same data for 28 clusters.  Interim zero points are based on a comparison of SMAC FP and SBF distances.}
\label{invnear}
\end{figure}

There are some details to be noted in the FP distances carried in {\it Cosmicflows-2}.  With both the SMAC and ENEARc samples we gave attention to the selection bias issue.  The authors of those studies addressed the problem in their own way.  We re-analyzed their data our way, analogous to the "inverse" TFR analysis.  Fits to cluster templates were made assuming errors on the velocity dispersion axis only.  Overall our results agree with SMAC and ENEAR literature results.  Figure~\ref{invnear} shows the agreement in the case of the ENEARc sample.  In a comparison between SMAC and ENEAR distances there is a slight reduction in scatter using values from our "inverse" fit analysis rather than the previously published values: the rms scatter between SMAC and ENEAR literature moduli for clusters in common is $\pm 0.222$ mag   while with our re-analysis the scatter is $\pm 0.189$ mag.   The distances that we report for SMAC and ENEAR galaxies are those determined by our new analysis.

In the case of EFAR we have not attempted a re-analysis.  Furthermore, the distances we use from this literature source are group averaged.  The clusters contributing to this study are at comparatively large distances and errors for individual galaxies are large.  In the {\it Cosmicflows-2} data catalog EFAR entries report distances that do not pertain just to the individual target but rather to their assigned group.

The connection between FP distances and TFR distances is shown in Figure 4 of \citet{2012ApJ...749..174C} with 32 clusters in common between the unified FP collection and SFI++ TFR and 11 clusters in common with CF2 TFR.  Again, correlations between the moduli are consistent with slope unity with distance.  Standard deviations with both SFI++ and CF2 comparisons are 0.04 mag.

\subsection{Integration of Type Ia Supernova Distances}

While we have argued that TRGB might be the gold standard for distances on small scales, it is evident that SNIa distances have become the gold standard on large scales.  The uncertainty in a single SNIa distance estimate is half the uncertainty of a TFR value.  A single SNIa measure is worth four TFR measures.  Even with relatively small samples there have already been interesting studies of large scale flow fields \citep{2011MNRAS.414..264C, 2011JCAP...04..015D, 2012MNRAS.420..447T}.

There has been tremendous activity in SNIa studies.  Observations are being made to relativistic redshifts that test cosmological models.  For our purposes we assembled SNIa distance measures from the literature \citep{2012ApJ...749..174C, 2012ApJ...758L..12S}.  Our compilation uses UNION2 \citep{2010ApJ...716..712A} as a backbone supplemented by four other studies of SNIa within $z=0.1$ \citep{2006ApJ...647..501P, 2007ApJ...659..122J, 2009ApJ...700.1097H, 2010AJ....139..120F}.  The merging of these samples is illustrated in Figure 1 of Courtois \& Tully.  Our dual purposes in those earlier papers were first to establish a bridge between the TFR and SNIa scales (the $I$ band TFR in Courtois \& Tully and the Spitzer [3.6] band in Sorce et al.) and second to use the ensuing SNIa calibration to arrive at a determination of the Hubble Constant in a redshift regime where peculiar velocities should be insignificant.

The full compilation of SNIa distances used in those earlier papers continue to help define H$_0$.  However, the CF2 compendium of distances truncates at $z=0.1$  Our other methodologies are limited to this regime.  Also, it is roughly at this redshift that there is a transition in the nature of SNIa surveys, from an approximation to all-sky for nearby SNIa to narrow angle (usually equatorial) deep imaging for distant SNIa.  We presently have a distillation of 306 SNIa distances within 30,000~\kms\ from the five sources identified above.  These have all been brought to the common CF2 zero point through comparisons demonstrated in Figure 5 of \citet{2012ApJ...749..174C}.

\subsection{Review of the Absolute Scale}

Relative distances are adequate for studies of peculiar velocities and the underlying density field but an accurate absolute determination of the Hubble Constant constrains the value of the equation of state for dark energy and the number of neutrino species \citep{2011ApJ...730..119R, 2012ApJ...758...24F}.   Our initial construction of the CF2 ladder of distances began with the HST Key Project scale based primarily on the Cepheid PLR and a distance modulus for the Large Magellanic Cloud of 18.50.  The details of this construction are found in the paper describing our calibration of the TFR at $I$ band \citep{2012ApJ...749...78T} and the subsequent paper on the extension of this calibration to SNIa \citep{2012ApJ...749..174C}.  The current discussion to this point has been based on the absolute scale described in those papers. 

The re-calibration of the TFR at the Spitzer [3.6] band \citep{2013arXiv1301.4833S} and its impact on the SNIa scale \citep{2012ApJ...758L..12S} provided an opportunity to re-evaluate issues concerning the absolute scale.  Thanks to trigonometric parallax observations with HST of Galactic Cepheids \citep{2007AJ....133.1810B}, studies of detached eclipsing binaries in the Large Magellanic Cloud \citep{1998ApJ...509L..21G, 2002ApJ...564..260F, 2002ApJ...574..771R, 2009ApJ...697..862P, 2013Natur.495...76P}, and mid-infrared observations of Cepheids in the LMC \citep{2012ApJ...758...24F} there is increased precision in the distance to the LMC.  Here we accept the modulus $18.48\pm0.03$ found by \citet{2012ApJ...758...24F}, slightly smaller than the HST key project fiducial distance although consistent within the assigned error.  This one change would have us decrease distances by 1\%.

The TFR distances that are such an important part of this compilation are all based on photometry at $I$ band.  Photometry for large samples based on Spitzer (and WISE) satellite observations in the mid-infrared are products for a future release of {\it Cosmicflows}.  In advance of that, there are several advantages with the mid-infrared photometry that can already be incorporated.  The most important advantage is the 1\% consistency of satellite photometry across the sky.  The $I$ band photometry has been acquired by many observers at a multitude of telescopes with different detectors and subtly different filters, north and south, over many observing seasons, and diverse not always well documented observing conditions.  The mid-infrared photometry has other advantages.  Reddening essentially disappears as a concern. The background (zodiacal light and distant galaxies) is at such a low level that almost all the light from a target is recorded in a short exposure.  The flux at $3.6\mu$m is dominated by old stars, presumed proxies for the mass.  Because of these advantages we trust the absolute calibration of the TFR at [3.6] more than we trust the $I$ band calibration.  It was determined that there is a small systematic difference between the [3.6] and $I$ band scales \citep{2012ApJ...758L..12S, 2013arXiv1301.4833S} though less than the error estimates.  With the HST key project LMC reference, distances are increased by 2\%.  In combination with the revision in the LMC distance, we end up with an increase in the CF2 scale by a factor 1.009, or 1\%.  

Accordingly, we make small adjustments as follows.  Distances that are directly based on Cepheid PLR observations are decreased in the modulus by 0.02 mag for consistency with the revised LMC distance; these include the Cepheid PLR measures themselves and the SBF measures.  Distances based on a Population II calibration (TRGB, RR Lyrae, Horizontal Branch) or geometric considerations (Eclipsing Binary, Maser) are unmodified.  Distances on large scales tied to our TFR calibration (TFR, FP, SNIa) are increased by 0.02 in the modulus to reflect the preferred mid-IR scale re-calibration.

\section{The Cosmicflows-2 Compendium of Distances}

In general, an individual galaxy has a distance by only one method.  Some methods have mutually exclusive samples, like TFR and FP.  There can be overlap between FP and SBF samples, even by design, but in this case we give preference to the SBF results.  The main exceptions to one distance method for one galaxy are for nearby objects.  Often in these cases the overlap is definitely by design.  Especially the Cepheid PLR results are used as a reference calibration for other techniques such as SNIa and TFR.  In these cases, the distances that we assign are coming from the calibrators; those drawn from the Cepheid PLR, TRGB, or the miscellaneous RR Lyrae, Horizontal Branch, Eclipsing Binary, or Maser contributions. There are overlapping distance estimates for 396 galaxies, 5\% of the total.

While distances from multiple techniques to individual galaxies are rare, multiple measures to different members of a group are common.  Group linkages are important for quite a few reasons.  Foremost is the $\sqrt{N}$ improvement that can accrue in distance uncertainties.  This same advantage is realized with systemic velocities, as we can average over all measures in a cluster (not just the galaxies with distances).  We also sum over all the $K$ band luminosity of the group \citep{2003AJ....125..525J, 2006AJ....131.1163S}.
In our compendium we report a group distance weighted over all estimates for group members, a mean group velocity and dispersion, and a total group $K$ luminosity.   Distance weights 1.0 are assigned if the source is Cepheid PLR, TRGB from our HST program, one of the miscellaneous contributions, or SNIa, weights 0.5 are assigned to TRGB values from the literature and SBF, weight 0.25 is given for a TF value, and 0.16 is given for FP.  These weights presume uncertainties of 10, 14, 20, and 25\% respectively in distances.

If our job is done correctly, mean distances to groups by different techniques will be consistent.  We will confirm this requirement with inter-comparisons between our three long-range methods: TFR, FP, and SNIa.  In addition, a group analysis provides a trap for bad data.  Outliers can be evaluated and, if appropriate, rejected.

{\it Cosmicflows-2} spans a huge range of conditions, from dense coverage that includes extreme dwarfs in the Local Group to the sparsest of sampling at $z \sim 0.1$.  A group analysis that encompasses this range is beyond the scope of this paper.  A proper group analysis must include many more galaxies than just those with measured distances.  For our present purposes we draw on two existing group catalogs, or perhaps it should be said three, as will be explained.

For the volume within 3000~\kms, we draw on the group catalog compiled by \citet{1987ApJ...321..280T} which was incorporated in the Nearby Galaxies Catalog \citep{1988ngc..book.....T}.  The original construction was rigorously defined through a dendogram built on estimators of mutual attraction between entities (luminosity/separation$^3$).  Over the years since, new-found galaxies have been added and the groups slightly re-arranged with access to new velocities and distances. The current version of the catalog is overdue for reconstruction but is still a good compilation of groups for the region within 40 Mpc.

The catalog that we use over the full range of CF2 is the compilation by  \citet{2011MNRAS.416.2840L} called 2MASSplusplus (2M++).  This group catalog was constructed with the 2MASS near-infrared extended source catalog \citep{2006AJ....131.1163S} as a base, with redshift information drawn from the all-sky 2MASS Redshift Survey to K=11.25 \citep{2005ASPC..329..135H},  the 6dF survey of the southern sky \citep{2009MNRAS.399..683J}, and the Sloan survey of the north galactic pole \citep{2009ApJS..182..543A}.  We use the group identifications from 2M++ as a secondary source, if not available in the Nearby Galaxies compilation or among the groups to be mentioned next.  In any event, we make use of the $K_s$ band luminosities made available by Lavaux \& Hudson in building group parameters.

The third source of group identifications is ad hoc.  Many contributions in the past with TFR and FP applications have involved observations of galaxies in clusters because of the advantage offered by averaging over many targets.  The SFI++ compilation and most of the FP programs are of this nature.  Many of the groups in question are Abell clusters \citep{1989ApJS...70....1A}.  We retain the use of clusters identified in those previous studies.  Frequently, the same groups or clusters have been observed in several programs, say, involving both TFR and FP.  Quite a few SNIa are found serendipitously to lie in these pre-defined groups. 

It deserves emphasis that our treatment of groups is preliminary.  Alternative group catalogs exist: locally \citep{2011MNRAS.412.2498M} and on large scales \citep{2007ApJ...655..790C}.  Ultimately, distance measurements will help inform group memberships.

\begin{figure}[h!]
\includegraphics[scale=0.42, angle=0]{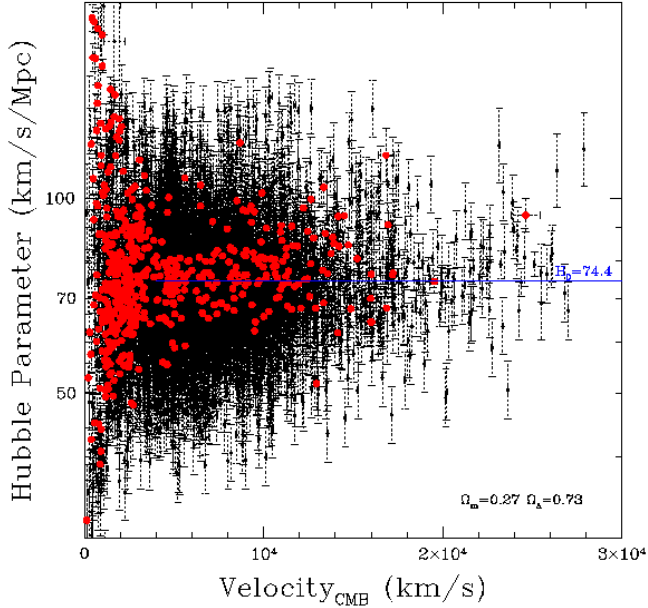}
\caption{Hubble parameter vs. velocity in the CMB frame for 534 groups (red) and 4690 single galaxies (black).  The weighted logarithmic mean for entities at $V_{cmb} > 4000$~\kms\ corresponding to H$_0 = 74.4$~\kmsMpc\ is given by the blue horizontal line. }
\label{hparam}
\end{figure}

In total, the present compilation has 1119 groups with at least two members contributing to a velocity average and 534 of these have at least two distance estimates.  The 8315 distance measures lie in 5224 entities, 3625 in groups and 4690 singles. Figure~\ref{hparam} shows values of the Hubble parameter, H$_i = V_{mod,i}/d_i$, for the 5224 entities, those in groups in red and singles black.  The velocity $V_{mod}$ includes relativistic corrections, small for these nearby galaxies, assume a cosmological model with $\Omega_m = 0.27$ and $\Omega_{\Lambda} = 0.73$.  
\begin{equation}
V_{mod}=c z [1+0.5(1-q_0)z-(1/6)(1-q_0-3q_0^2+1)z^2]
\end{equation}
where $z$ is redshift in the CMB frame and $q_0 = 0.5(\Omega_m-2\Omega_{\Lambda})$.
The weighted fit in the logarithm to the Hubble parameter for 3996 groups and singles with $V_{cmb} > 4000$~\kms\ leads to $<{\rm H}_i> = 74.4 \pm 0.2$~\kmsMpc.  The error is the statistical standard deviation with the very large sample.  This mean Hubble value found for constituents in the redshift interval $0.013 < z < 0.1$ is consistent with the mid-IR calibration value found by fits to SNIa in the range $0.03 < z < 0.5$ \citep{2012ApJ...758L..12S}  of H$_0 = 75.2 \pm 3.0$~\kmsMpc, where in this latter case the error includes random and systematic uncertainties. 

\begin{figure}[h!]
\includegraphics[scale=0.52, angle=0]{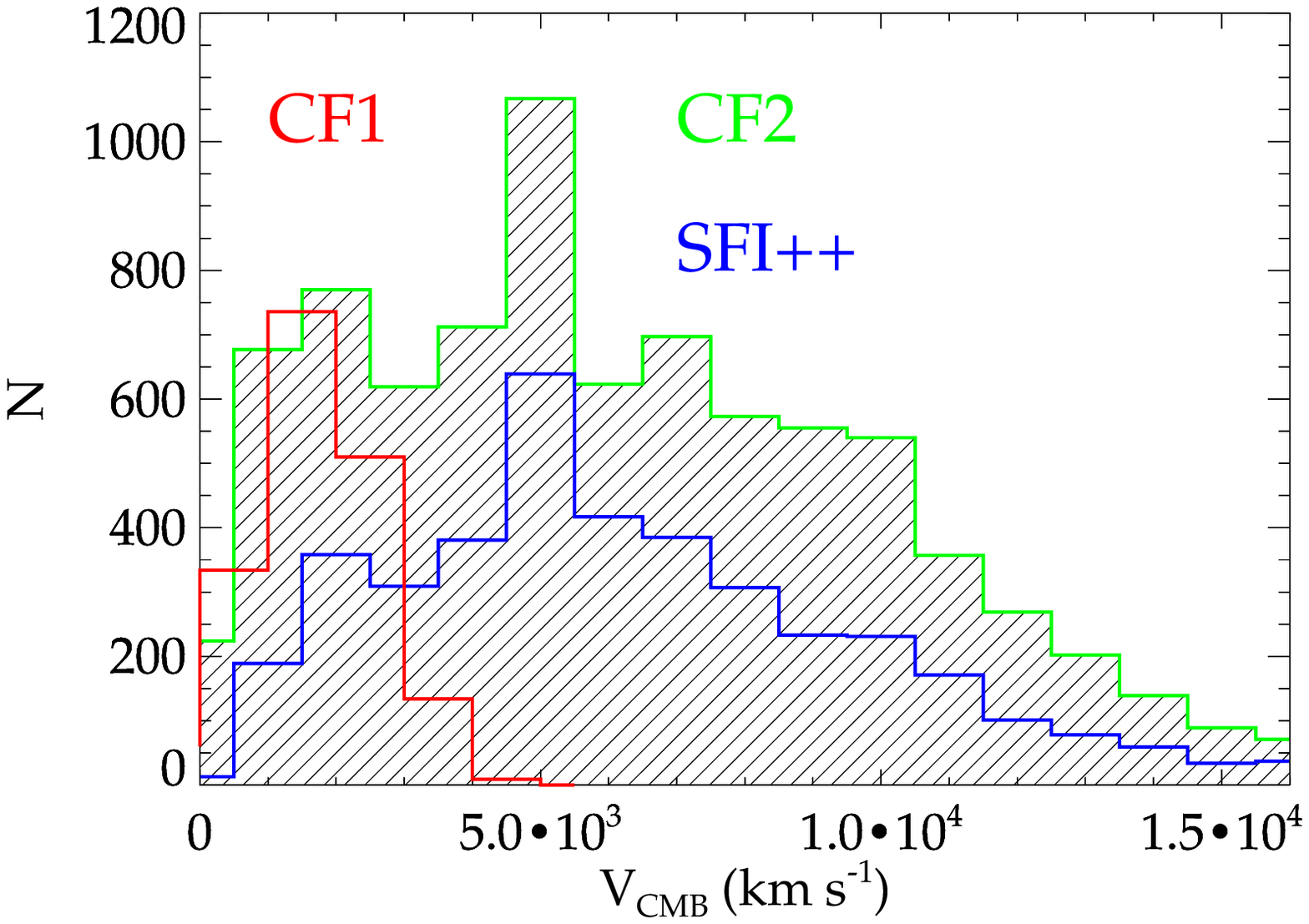}
\includegraphics[scale=0.52, angle=0]{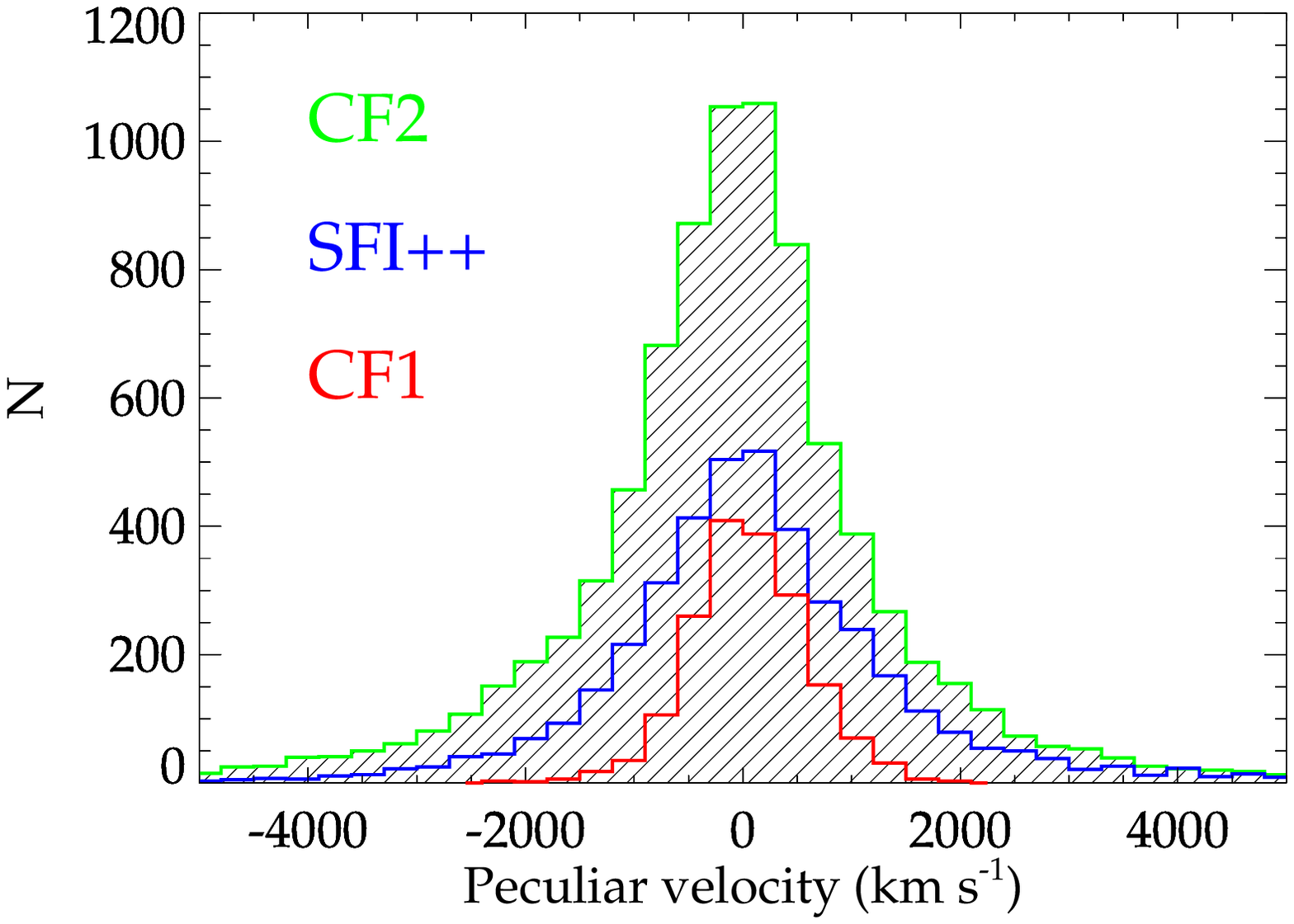}
\caption{Histograms of systemic velocities in the CMB frame (top) and peculiar velocities (bottom) for the {\it Cosmicflows-2} sample (shaded, green outline) and, for comparison, equivalent histograms for the {\it Cosmicflows-1} sample (red outline) and the SFI++ sample (blue outline).}
\label{histv}
\end{figure}

\begin{figure}[h!]
\includegraphics[scale=0.81, angle=0]{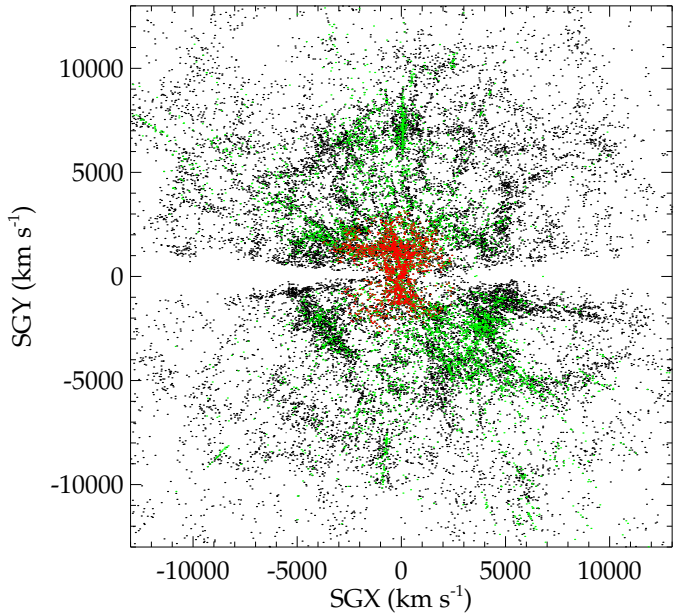}
\caption{Projection of the distribution of galaxies on the supergalactic equator.  All galaxies from the 2MASS $K<11.75$ redshift survey with SGZ $< \pm4000$~\kms\ in black, galaxies with distance measures in {\it Cosmicflows-1} in red, and galaxies with distance measures in {\it Cosmicflows-2} in green.  The plane of our Galaxy lies at SGY=0 and causes the wedges of evident incompletion.}
\label{xscz}
\end{figure}

Figure~\ref{histv} provides a summary of gross properties of the sample.  The top panel illustrates the distribution of the full sample as a function of systemic velocity and the bottom panel shows the distribution of peculiar velocities, assuming H$_0 = 74.4$~\kmsMpc, for the same full sample.  For comparison, the same distributions are shown for the {\it Cosmicflows-1} and SFI++ compilations of distances.  CF1 has comparable completion nearby but is restricted in depth.  SF++ explores a comparable range of distances but with lower density.  The distribution in a slice of space is seen in  Figure~\ref{xscz}.  In this plot, black points are drawn from a redshift survey, the red overplots show the coverage already available with CF1, and the green overplots show the coverage now available with CF2.

\subsection{An Error Bias in Peculiar Velocities}

\begin{figure}[h!]
\includegraphics[scale=0.68, angle=0]{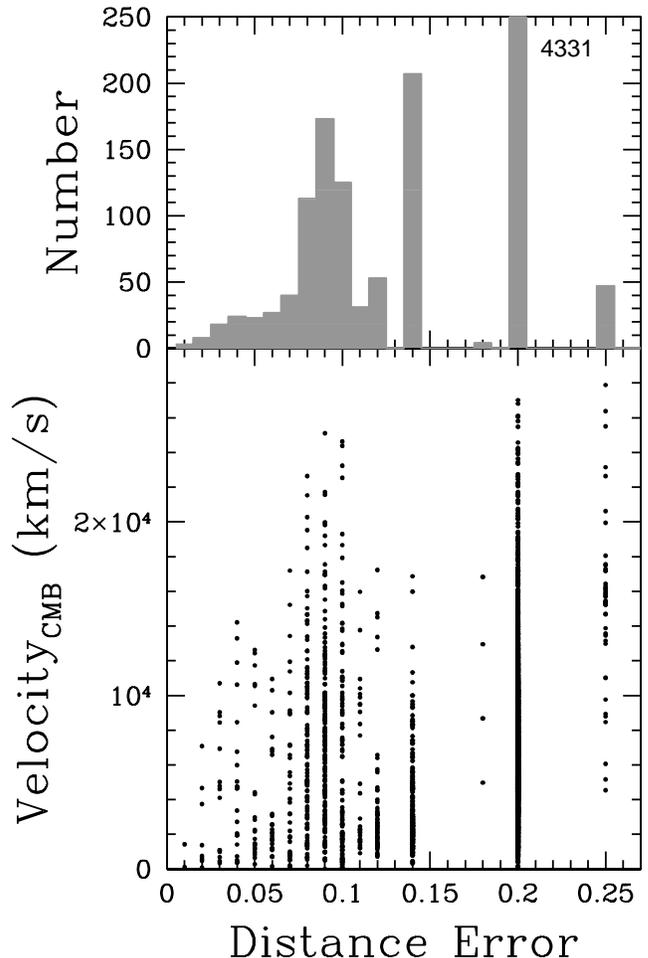}
\caption{Error assignments for the 5224 groups and single galaxies with distance estimates. {\it Top:} Histogram of fractional distance error assignments.  There are 4331 singles with TFR measures assigned uncertainties of 20\%. {\it Bottom:} The individual error assignments are plotted as a function of systemic velocity.}
\label{ed}
\end{figure}

The entire sample is a mix of distance measures with uneven quality.  Where possible, group assignments and subsequent averaging considerably reduce both distance and velocity errors.  Uncertainty estimates are summarized in Figure~\ref{ed} for the 5224 distinct entities after grouping.  The top panel gives a histogram of the uncertainty estimates.  The distribution is strongly bimodal.  One peak at distance errors $8-10\%$ results from the input of precision estimators such as Cepheid PLR, TRGB, and SNIa and from groups with many contributions (the entities with the lowest error estimates are either in the proximity of the Local Group or, such as the Virgo Cluster, with many contributions $-$ in such cases systematic errors dominate these statistical error estimates).  The other peak in the bimodal error distribution is at 20\%, the error assigned to individual TFR distances.  In the figure, this contribution goes far off scale.  The lower panel illustrates the distribution of the error estimates with systemic velocity.

As is well known, if percentage errors are roughly a constant as a function of distance with a given methodology then errors in assigned peculiar velocity grow linearly with distance, rapidly becoming much larger than intrinsic peculiar velocities.  The problem is aggravated by an asymmetry in the error-induced peculiar velocities.  The errors are symmetric in measurements of distance modulus, a logarithmic quantity.  However there is a resultant skewness in distance errors.  As a thought experiment, consider two galaxies that are intrinsically at 100 Mpc and have no peculiar velocity with H$_0=75$, whence they are observed at 7500~\kms.  Suppose that rather than observed to be at their correct distance modulus of 35.0, one has a distance modulus error of $2\sigma$ with a TFR measurement that brings it forward to 34.2, while the other has a comparable error that takes it back to 35.8.  The corresponding distances are 69 and 145 Mpc and the implied peculiar velocities ($V_{pec} = V_{mod} - {\rm H}_0 d$) are +2311 and -3338 \kms\ respectively. It follows that there is a skewness in peculiar velocities, with the tail to negative velocities more extensive than the tail to positive velocities.

\begin{figure}[h!]
\includegraphics[scale=0.44, angle=0]{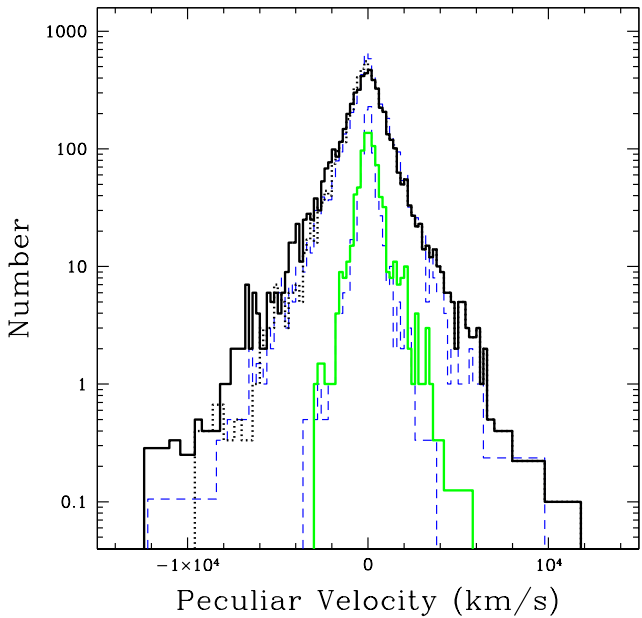}
\caption{Histogram of peculiar velocities: all groups and singles (outer black histogram) and only entities with fractional distance error estimates $\le 0.14$ (inner green histogram).  The dashed blue histograms were built from the same samples, assuming deviations from Hubble distances at each $V_{mod}$ $\pm$ random gaussian deviations based on the associated assumed error in the distance modulus.  The black dotted histogram at negative peculiar velocities shows the distribution after the adjustment for the distance error bias.}
\label{histNVp}
\end{figure}

This phenomenon is observed in the peculiar velocities inferred from CF2 distance measurements.  The effect can already be seen in the lower panel of Figure~\ref{histv} but is more clearly seen with the Log N scale used in the histogram of Figure~\ref{histNVp}.  Here, the black outer histogram is built with the entire grouped sample assuming H$_0 = 74.4$~\kmsMpc\ while the inner green histogram reflects only the sub-sample with distance error estimates $\le 14\%$.

That the skewness toward negative velocities evident in both the full sample and sub-sample are due to the error bias is demonstrated by the good matches provided by the dashed histograms.  These secondary histograms are generated from a mock file that duplicates the systemic velocities and error estimates of the real data.  Distances are assigned to the mock galaxies by giving them random deviations from the Hubble value, drawing from a gaussian distribution in the modulus with the gaussian dispersion given by the error estimate in the particular case.

The mock experiment guides an adjustment that effectively removes the error bias to peculiar velocities.  Positive peculiar velocities are untouched but negative peculiar velocities are shrunk by a small factor controlled by the fractional distance uncertainty $e_d$ as described by the following equation.
\begin{equation}
V_{adj} = V_{pec} [0.77 + 0.23 e^{-0.01(e_d V_{mod})}]
\end{equation}
The adjusted velocity $V_{adj}$ approaches $0.77 V_{pec}$ if the product $e_d V_{mod}$ is large and approaches $V_{pec}$ if that product is small.
The adjustment is justified by the reality that errors strongly dominate the peculiar velocity signal in the regime of significant modifications. In the tables to be discussed, both raw and adjusted peculiar velocities are made available.

\begin{figure}[h!]
\includegraphics[scale=0.44, angle=0]{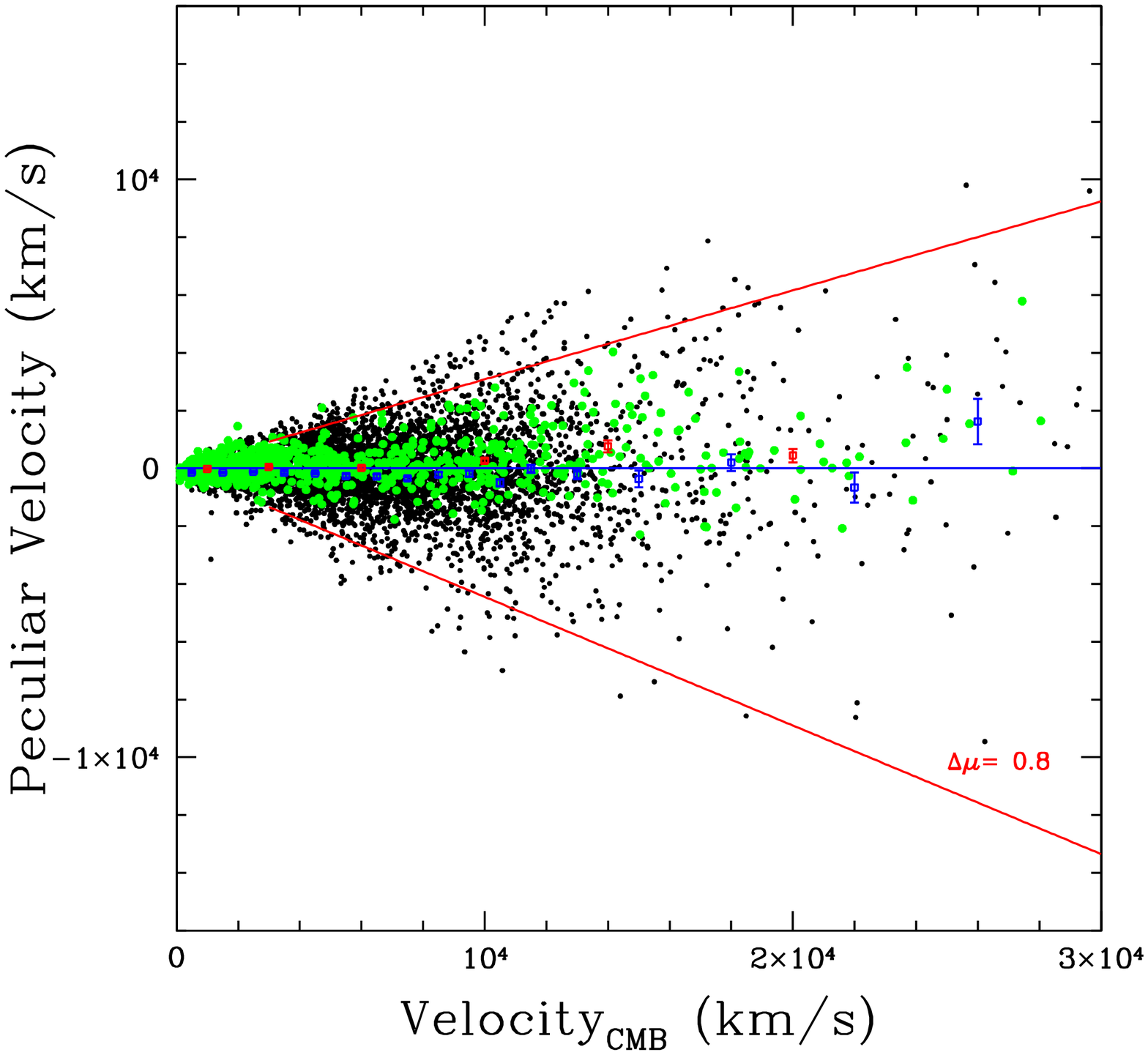}
\includegraphics[scale=0.44, angle=0]{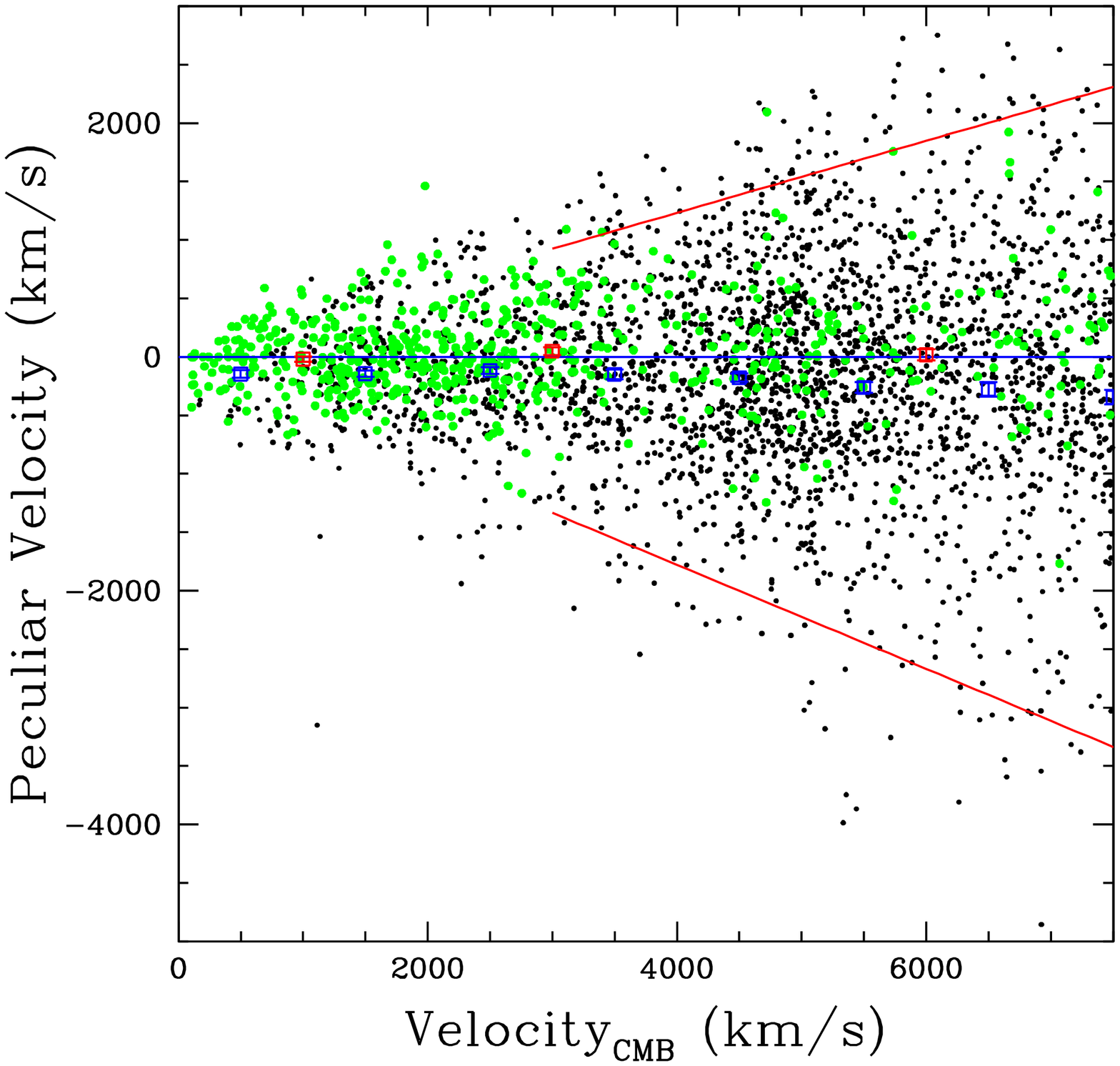}
\caption{Peculiar velocities, with H$_0=74.4$ \kmsMpc, vs. systemic velocity.  The lower panel is an expansion of the scale of the upper panel.  Green points: error assignments $\le 14\%$ (red boxes with error bars are binned averages), black points: error assignments $> 14\%$ (blue boxes with error bars are binned averages). Negative peculiar velocities have been adjusted to negate the error bias to to peculiar velocities.
Red solid lines: loci of erroneous peculiar value estimates that would result from a $2\sigma$ error in a TFR distance estimate.}
\label{pecv}
\end{figure}

The increase in the amplitude of measured peculiar velocities with redshift because of errrors is evident in the two panels of Figure~\ref{pecv}, the lower panel simply an enlargement of the crowded region in the top panel.  The black points in these displays show the individual peculiar velocities as a function of systemic velocity for the full grouped sample while the green points emphasize the sub-sample with fractional errors $\le0.14$.  The blue and red squares with error bars (some errors too small to be seen) are averages in systemic velocity bins.    The effect of distance error bias is easily seen in the peculiar velocity asymmetries of the guide lines in this figure.  The solid red lines show the peculiar velocity loci with distance modulus errors $\pm0.8$ mag, a $2\sigma$ error with a TFR measurement.  In the figure, the black and green points have been plotted after applying the distance error adjustment to negative peculiar velocities, with the consequence that the bias toward larger amplitude negative velocities is removed.  It can be seen in Figure 17 that extreme peculiar velocities are clipped.  In all, 138 galaxy distance measurements are rejected either because they have greater than $3\sigma$ excursions from a group mean or because of extreme Hubble parameter excursions.  In two thirds of these cases the reason for a bad distance measurement was evident on close inspection, attributable to a bad inclination, strange morphology, confusion, or interaction with a neighbor.  The number of rejections is 1.7\% of the total sample.

\subsection{The Catalogs}

The {\it Cosmicflows-2} data are made available in two tables: one providing an entry for every galaxy with a distance and the other condensed to an entry for each separate group, including groups of one galaxy.  Table 1 is the complete catalog, made available in its entirety on-line, and also available at EDD, the Extragalactic Distance Database, with versions that might be updated.  The following is a description of the 8315 current entries in Table 1.

\noindent
{\it Col. 1:} Principal Galaxies Catalog (LEDA) number.

\noindent
{\it Col. 2-4:} Distance, distance modulus, and fractional distance error for the galaxy.  Weighted average values are given if there are multiple sources.

\noindent
{\it Col. 5-12:} Codes indicating source of distance: C=Cepheid PLR; T=TRGB from this program; L=TRGB from literature; M=miscellaneous (RR Lyr, Horizontal Branch, Eclipsing Binary, Maser); S=SBF; N=SNIa; H=TFR; F=FP.

\noindent
{\it Col. 13-18:} Coordinates, successively celestial (J2000), Galactic, and supergalactic.

\noindent
{\it Col. 19:} Morphological type in the RC3 numeric code.

\noindent
{\it Col. 20:} $B$ band reddening from Schlegel et al. (1998).

\noindent
{\it Col. 21-22:} Magnitudes at $B$ and $K_s$ bands from RC3 and 2MASS respectively.

\noindent
{\it Col. 23-27:} Velocities in the successive reference frames: helio, Galactic, Local Sheet, CMB, and CMB adjusted for cosmological effects with $\Omega_{m}=0.27$ and $\Omega_{\Lambda}=0.73$.

\noindent
{\it Col. 28:} Common name.

The following columns pertain to the group associated with the individual galaxy.

\noindent
{\it Col. 29:} Bookkeeping group numer; preferred group identification.

\noindent
{\it Col. 30:} 2M++ group ID (Lavaux \& Hudson 2011); alternate group identification. 

\noindent
{\it Col. 31:} Number of galaxies with measured distances in group.

\noindent
{\it Col. 32-34:} Weighted average distance, distance modulus, and fractional distance error of group.

\noindent
{\it Col. 35:} Number of galaxies with known positions and velocities in group.

\noindent
{\it Col. 36-39:} Galactic and supergalactic coordinates of group.

\noindent
{\it Col. 40:} Mean morphological type of group members.

\noindent
{\it Col. 41-42:} Summed $B$ and $K_s$ magnitudes for group.

\noindent
{\it Col. 43-47:} Mean group velocity; respectively helio, Galactic, Local Sheet, CMB, and CMB adjusted for cosmological effects with $\Omega_{m}=0.27$ and $\Omega_{\Lambda}=0.73$.

\noindent
{\it Col. 48:} RMS group velocity dispersion.

\noindent
{\it Col. 49:} Alternate names for group or cluster.

Table 2 contains 5224 group entries, including 4690 of these as singles.  The column entries track entries 28-47 in Table 1 and in addition include 3 more velocity related parameters. The following information is included.

\noindent
{\it Col. 1:} Number of galaxies with measured distances in group.

\noindent
{\it Col. 2-4:} Weighted average distance, distance modulus, and fractional distance error of group.

\noindent
{\it Col. 5:} Number of galaxies with known positions and velocities in group.

\noindent
{\it Col. 6-9:} Galactic and supergalactic coordinates of group.

\noindent
{\it Col. 10:} Mean morphological type of group members.

\noindent
{\it Col. 11-12:} Summed $B$ and $K_s$ magnitudes for group.

\noindent
{\it Col. 13-16:} Mean group velocity; respectively helio, Galactic, Local Sheet, CMB.

\noindent
{\it Col. 17:} $V_{mod}$, group velocity with adjustment for cosmological model ($\Omega_m=0.27$, flat topology) as given by Eq. 14.

\noindent
{\it Col. 18:} RMS group velocity dispersion.

\noindent
{\it Col. 19-20:} Value in column 19 is peculiar velocity $= V_{mod} - {\rm H}_0 d$ assuming H$_0 = 74.4$~\kmsMpc.   Value in column 20 is peculiar velocity adjusted for the distance measurement error bias (only differs from value in previous column if negative).

\noindent
{\it Col. 21:} Bookkeeping group number; preferred group identification.

\noindent
{\it Col. 22:} 2M++ group ID (Lavaux \& Hudson 2011); alternate group identification. 

\noindent
{\it Col. 23-24:} Alternate names for group or cluster; PGC number identifies brightest galaxy in group.

\section{Discussion}

Arguably more interesting than galaxy distances are galaxy peculiar velocities.  However, metric errors in distance and hence in peculiar velocities tend to increase linearly with distance.  It is only very nearby that individual measured peculiar velocities dominate over errors.  Averaging over neighbors is increasingly necessary with distance.  Also, although individual distances may be unbiased, the homogeneous and non-homogeneous Malmquist effects and error bias can generate spurious artifacts in velocity fields.  These issues must be addressed if velocities are used to infer the distribution of matter.  For example, the Malmquist effects have dramatically reduced impact in an analysis carried out in redshift space rather than physical space.  

It is beyond the scope of this paper to dwell on these issues.  We are content here to offer a couple of teases, one local and another on a large scale.  Nearby, thanks to the HST TRGB program, there is increasingly detailed information on galaxy clustering and motions.  Within 10 Mpc of our position, group assignments are unambiguous and peculiar velocities dominate errors.  Already with {\it Cosmicflows-1} there was dense local coverage and a major conclusion (Tully et al. 2008) was that all the galaxies in the Local Sheet are moving coherently away from the Local Void (at 260 \kms) and toward the Virgo Cluster (at 185 \kms).  However at the time of that earlier study there was limited information on the gradients of the velocity flows, either vertically with respect to the Local Sheet along the flow out of the Local Void or within the Local Sheet where a shear should develop as one approaches the Virgo Cluster.  

\begin{figure}[h!]
\includegraphics[scale=1., angle=0]{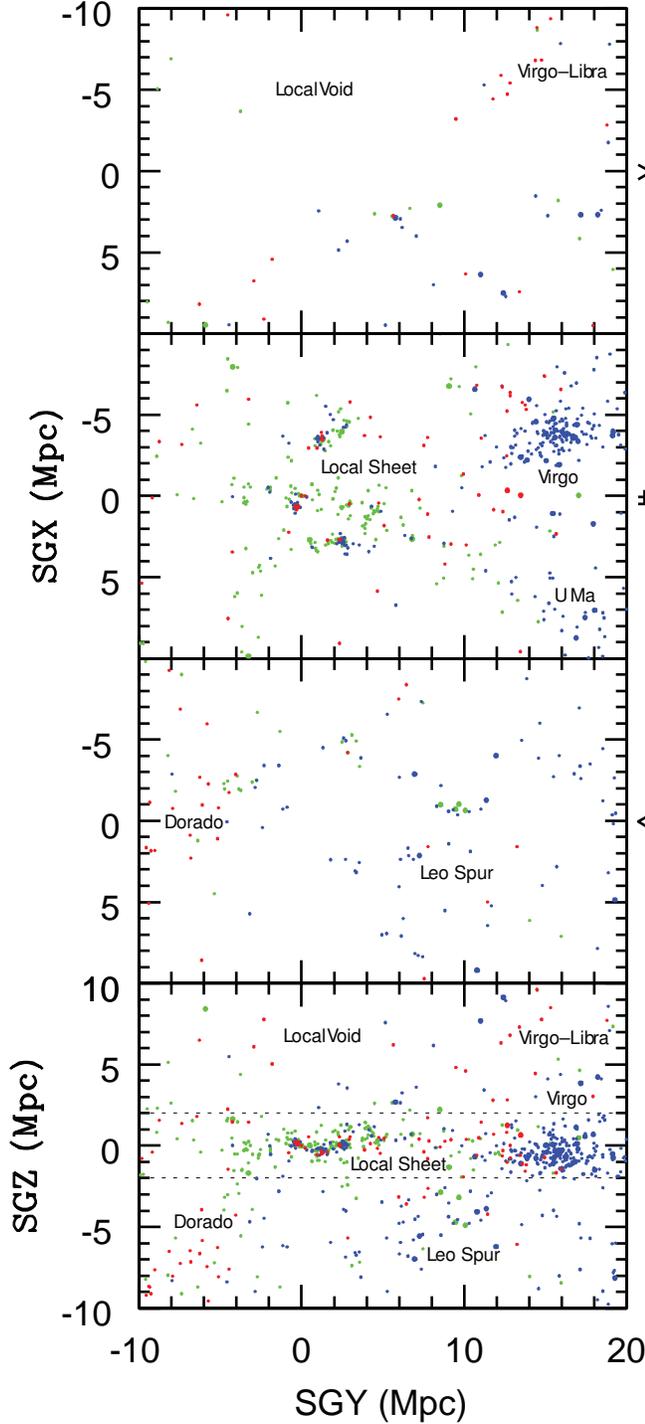}
\caption{Nearby peculiar velocities. The Milky Way is at the origin in each frame.  The view in the bottom panel is edge-on to the plane of the Local Sheet while this structure is viewed face-on in the panel 2nd from the top.  The top and 3rd panels show regions above and below the Local Sheet respectively.  Galaxies with peculiar velocities less than $\pm100$~km/s\ are represented in green.  Galaxies with peculiar velocities greater than 100~\kms\ are red and those at less than $-100$~\kms\ are blue.  Prominent features are labeled.}
\label{xyzloc}
\end{figure}

Recent TRGB observations are beginning to clarify these situations.  
Figure~\ref{xyzloc} shows plots of where galaxies lie that presently have have distance determinations.  The lower panel presents an edge-on view of the Local Sheet extending to the Virgo Cluster at the extreme right.  Three zones are distinguished by the horizontal dashed lines at $\pm 2$ Mpc.  The top three panels give polar views of the top, equatorial, and lower zones respectively.  Our Galaxy is at (0,0,0).  Peculiar velocities (residuals from Hubble expansion with H$_0=74.4$~\kmsMpc) are coded with three colors: green for peculiar motions less than $\pm 100$~\kms, red for those greater than 100~\kms, and blue for those more negative than $-100$~\kms.  The main results from the CF1 study are recovered. Peculiar velocities within the Local Sheet are small (most nearby galaxies are colored green) but the Local Sheet has a bulk motion (galaxies below and to the right of the Local Sheet in the lower panel are colored blue because the Local Sheet is moving down, away from the Local Void, and to the right, toward the Virgo Cluster).  

Major features can be followed across the panels.  The entities labeled Virgo-Libra and Dorado, borrowing terminology from the Nearby Galaxies Atlas (Tully \& Fisher 1987), are predominantly colored red.  These features are closer than anticipated from their redshifts and moving away in co-moving space.  Both features are other parts of the wall bounding the Local Void.  

\begin{figure}[h!]
\includegraphics[scale=0.44, angle=0]{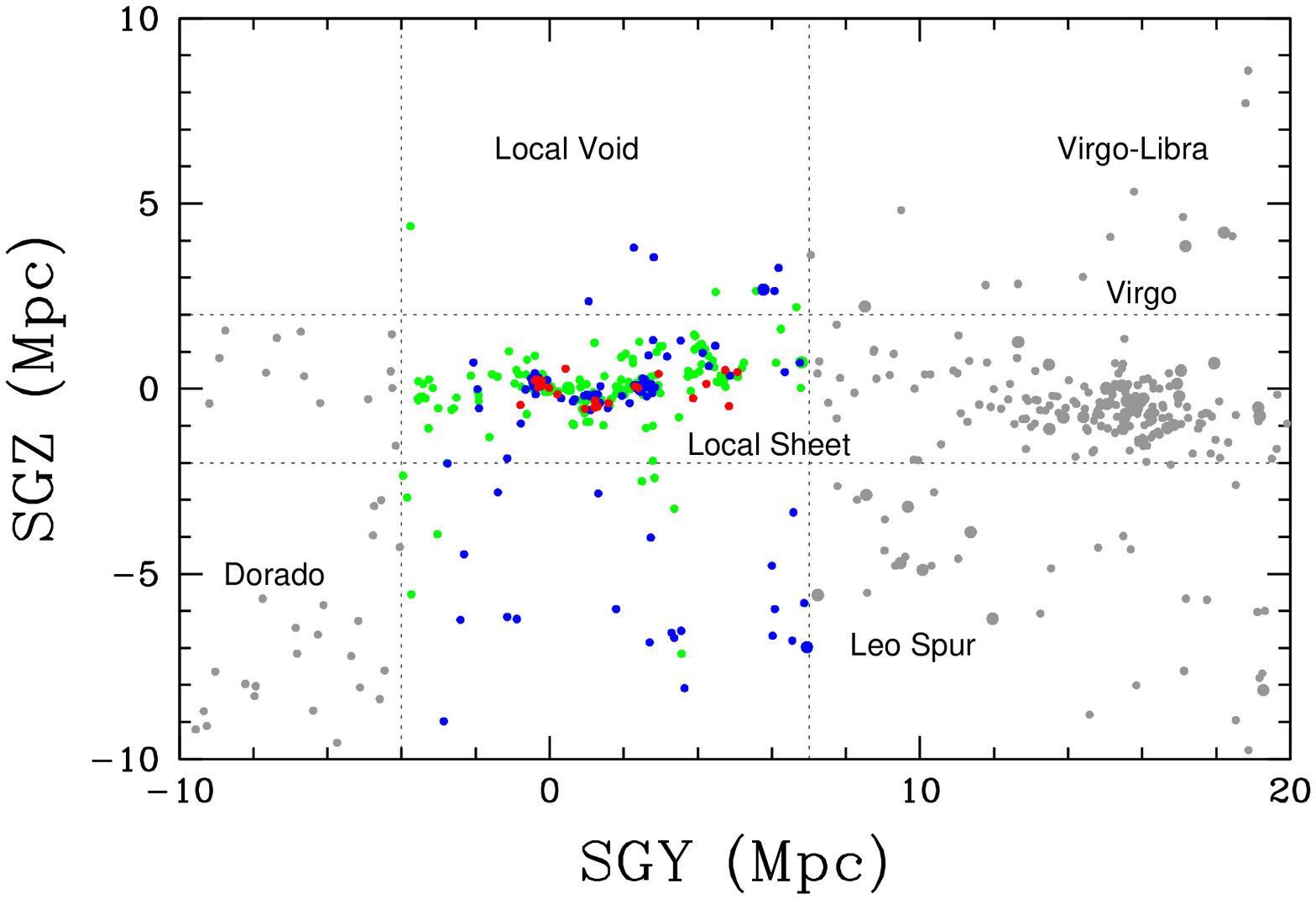}
\includegraphics[scale=0.44, angle=0]{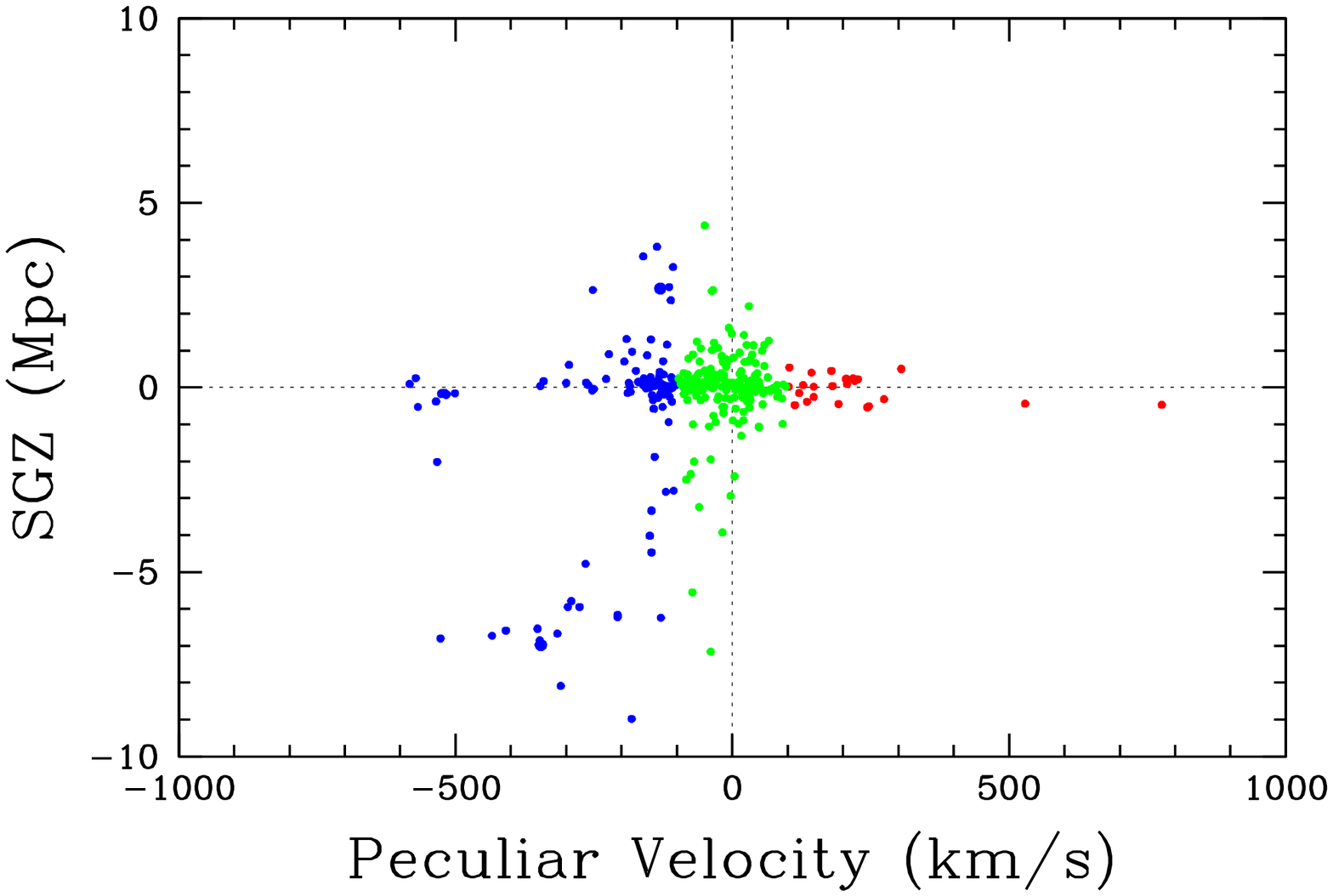}
\caption{Peculiar velocities vertically through the Local Sheet.  The top panel illustrates the region being considered.  The depth is restricted to $-7<SGX<+4$~Mpc.  The bottom panel plots the peculiar velocities of galaxies in this region as a function of distance above and below the equatorial plane of the Local Sheet.  Almost all galaxies both above and below the Local Sheet have negative peculiar velocities.}
\label{drillz}
\end{figure}

Figure~\ref{drillz} gives detail to the expansion of the Local Void where we can study it best, immediately in our vicinity.  The top panel isolates the region of interest.  The bottom panel shows the distribution of peculiar velocities in this region as a function of supergalactic latitude.  Essentially all galaxies above and below the equatorial plane have negative peculiar velocities.  This pattern is the signature of void expansion.  The Local Sheet is moving downward, catching up to objects below the plane.  Galaxies above the plane are moving faster still, catching up to us as they evacuate the void. 

\begin{figure}[h!]
\includegraphics[scale=0.44, angle=0]{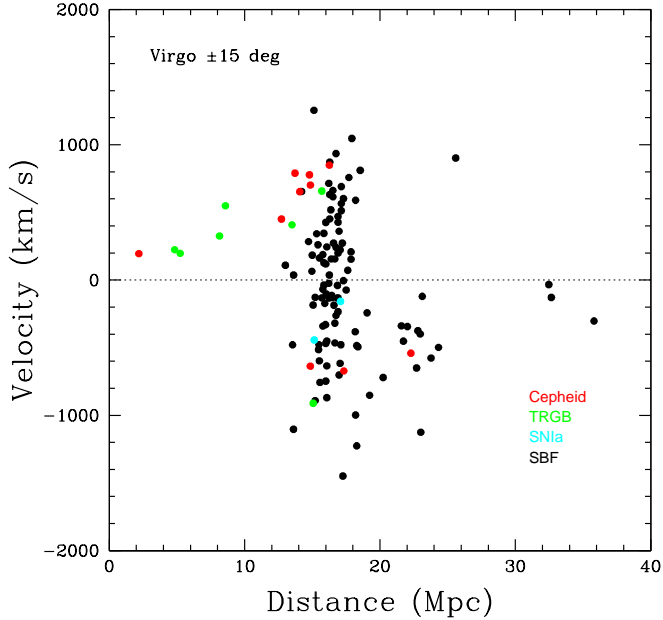}
\caption{Co-moving velocities as a function of distance in a $15^{\circ}$ radius cone centered on the Virgo Cluster. Hubble expansion with $d_{virgo}=16.37$~Mpc and H$_0=74.4$~\kmsMpc\ has been subtracted. Only distances with low errors are plotted.  Colors identify distance methodologies.}
\label{virZwave}
\end{figure}

Returning to Figure~\ref{xyzloc}, it can be seen in the panel second from the top, the face-on view of the Local Sheet, that there are quite a few objects colored red between the green of the Local Sheet and the blue of the Virgo Cluster.  These are galaxies experiencing the cluster infall shear.  The flow pattern is demonstrated in Figure~\ref{virZwave}.  The galaxies represented here lie within a $15^{\circ}$ cone centered on the cluster.  Only galaxies with high quality distance measures are represented.  Hubble expansion has been subtracted from velocities.  Colors identify the techniques used to derive individual distances.  The $Z$-wave pattern of infall is becoming increasingly well defined with the addition of new data.

\begin{figure}[h!]
\includegraphics[scale=0.44, angle=0]{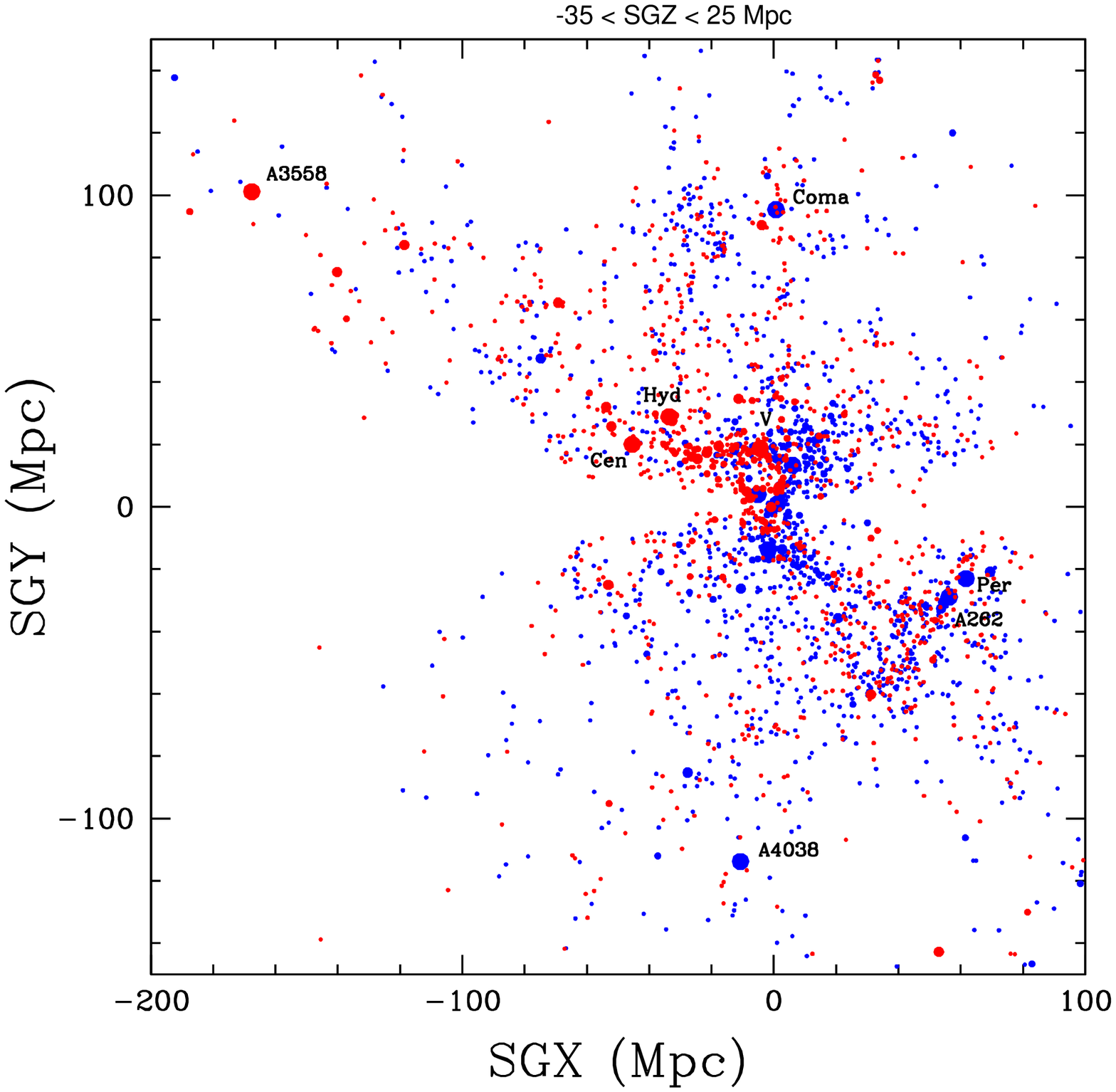}
\includegraphics[scale=0.44, angle=0]{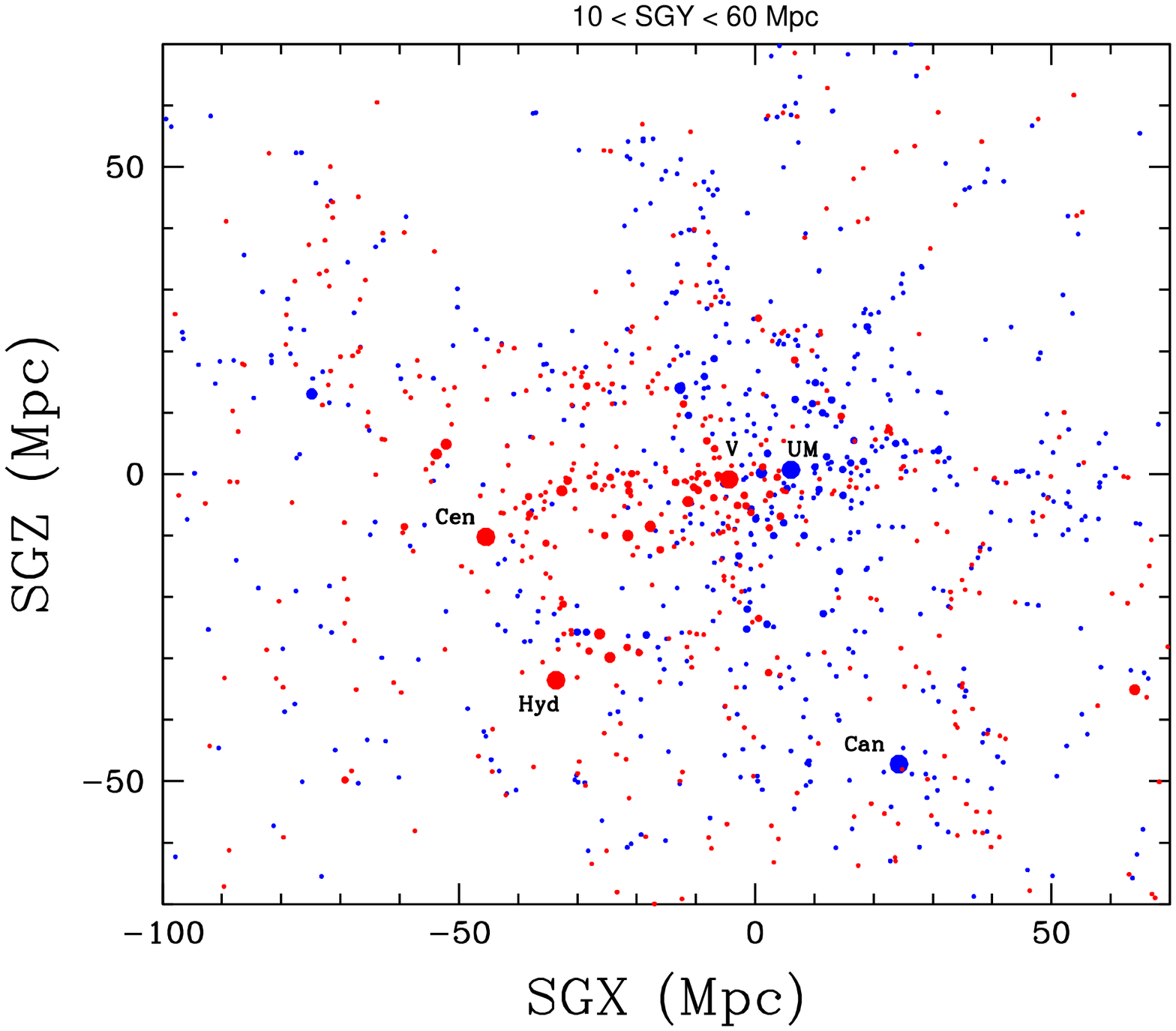}
\caption{Two views of peculiar velocities on large scales.  Galaxies with measured distances are located in redshift space and colored red if peculiar velocities are positive or blue if negative.  Clusters with more than 30 measures are given the largest symbols and identified.  The top panel presents a slice 60 Mpc thick on the supergalactic equator.  The zone of obscuration creates a data gap in wedges on the mid-plane.  The lower left quadrant corresponds to the celestial south where CF2 provides poor coverage beyond $\sim 40$ Mpc.  In the lower panel, the view is edge-on to the supergalactic equatorial plane and the slice is 70 Mpc thick offset to positive SGY to include the main body of local structure.  The scale is amplified in the lower panel.  In both panels, there are extended regions where one color predominates over the other, indicative of large scale systematic flows.}
\label{vlss}
\end{figure}

This discussion will be brought to a close with an ever so brief look at peculiar velocities on large scales.  Two scenes are presented in Figure~\ref{vlss}.    The top panel is a polar view of peculiar velocities with structure in the supergalactic equatorial plane and the lower panel gives an edge-on view of the main body of the structure.  Our Galaxy is at the origin.  In the top view, the objects are mostly red in the upper left quadrant, indicative of motion away from us, and mostly blue in the lower and right quadrants, indicative of motion toward us.  This pattern is a manifestation of a flow with shear toward the upper left, the familiar flow toward the "Great Attractor"  \citep{1987ApJ...313L..37D} in the vicinity of the clusters labeled Cen (Centaurus) and Hyd (Hydra).  The preponderance of red symbols continues to the upper left as far as Abell 3558 in the Shapley Supercluster \citep{1989Natur.342..251R} with a smattering of blue at intermediate distances hinting at a dip in the flow.    Manifestations of the same pattern are seen in the lower panel.

\begin{figure}[h!]
\includegraphics[scale=0.44, angle=0]{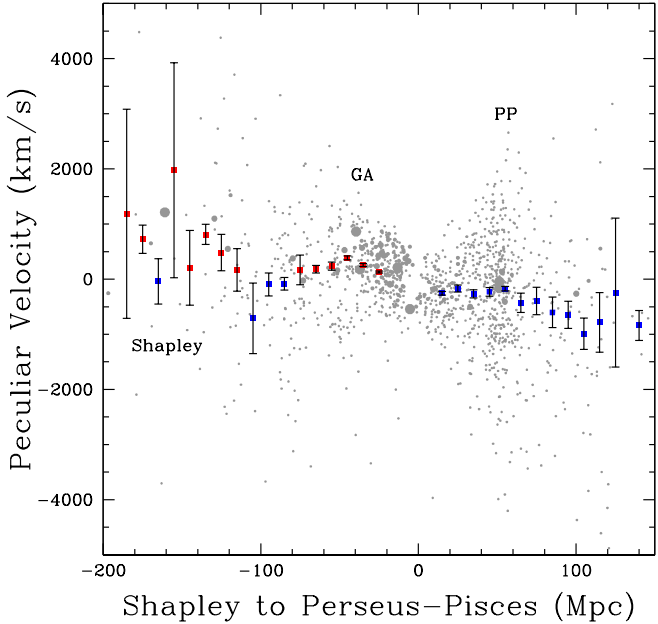}
\caption{Peculiar velocities in a band running from the Shapley Supercluster to the Perseus-Pisces filament.  Individual objects are plotted grey and averages in 10 Mpc bins are given color and error bars, with positive peculiar velocities in red and negative in blue.  The distance ranges of Shapley, Great Attractor (GA), and Perseus-Pisces (PP) are identified.}
\label{cord}
\end{figure}

The flow pattern running from 4 o'clock to 10 o'clock in the top panel of Figure~\ref{vlss} can be examined in greater detail in Figure~\ref{cord}.  This figure shows peculiar velocities in the lower-right and upper-left quadrants of Fig.~\ref{vlss}, top, that lie in a band $\pm 20$ Mpc wide in SGY running through the origin and tilted $34^{\circ}$ from the SGX axis.  Velocity values are averaged in 10 Mpc intervals along the axis of the band.  There is an evident gradient of increasingly positive peculiar velocities proceeding from right to left.  Between the Centaurus-Hydra (Great Attractor) region and the Shapley region, however, there is a dip toward negative velocities that is reasonably convincing.   Back-side infall into the Centaurus-Hydra region would create such a signature.  

\section{Summary}

{\it Cosmicflows-2} is a compilation of 8315 distances within 5224 entities: 3625 of these distances in 534 groups and 4690 singles.  The full compilation includes entries for all 8315 galaxies and provides both an individual distance and velocity for the entry and an averaged group distance and group velocity in instances of a group assignment.  The TFR contributes 5998 measures and FP contributes a non-overlapping 1508.  Numerically these sources dominate but they suffer the largest uncertainties.  Candidates have frequently been selected because they lie in clusters (particularly FP) and group averaging results in significant improvements.

Although TFR and FP contributions constitute the bulk of the distance estimates, roughly 1000 come from other methods with individually higher accuracy, including 60 Cepheid PLR, 297 TRGB from our program, 133 TRGB from the literature, 382 SBF, 306 SNIa, and 31 miscellaneous (Horizontal Branch, RR Lyr, Eclipsing Binary, or Maser).  The Cepheid, TRGB, and miscellaneous methods serve as the building blocks for the other procedures.  They provide the zero point scaling for TFR and SBF distances.  These in turn provide the scaling for FP distances through clusters observed in common.  SNIa are brought to a common scale through either individual galaxy matches or cluster matches.  The final scale reflects refinements to the LMC distance \citep{2012ApJ...758...24F, 2013Natur.495...76P} and a TFR calibration with mid-IR photometry with Spitzer Space Telescope \citep{2012ApJ...758L..12S}.  The weighted value for the Hubble Constant found for contributions at $V_{cmb} > 4000$~\kms\ is H$_0 = 74.4$~\kmsMpc.  The statistical error with such a large sample is small and totally dominated by systematics of the calibration.  As an estimate of the uncertainty on H$_0$ we accept the calibration error budget \citep{2012ApJ...749..174C, 2012ApJ...758L..12S} of $\pm3.0$~\kmsMpc.
While the value for H$_0$ reported here of $74.4\pm3.0$ is consistent with the recent directly observed values of $73.8\pm2.4$ \citep{2011ApJ...730..119R} and $74.3\pm2.6$ \citep{2012ApJ...758...24F} it is at odds with the Planck 2013 indirect value of $67.3\pm1.2$~\kmsMpc \citep{2013arXiv1303.5076P}.

Although {\it Cosmicflows-2} provides a value for the Hubble Constant that should be competitive with the best of alternative derivations, the greatest interest is with the information it provides regarding departures from Hubble expansion.
A critical point to appreciate is that the computation of peculiar velocities requires a value of H$_0$ compatible with the zero point and scale of the observations, irrespective of whether it is the correct value.  Assuming the Planck value, for example, would lead to systematic monopole outflow with the Cosmicflows-2 distances.  If the Planck result is correct it implies a scale error in the present data set at the level of 10\%, a circumstance we consider unlikely at the $2.5 \sigma$ level.

On large scales measurement errors to distances, hence peculiar velocities, dominate intrinsic motions in individual cases so only averaged values will have meaning and only the most elementary of bulk flows will be determined with confidence.  Very nearby, though, the situation is quite different.  There is an increasingly clear pattern emerging.  Galaxies within filaments have very small random motions with respect to their neighbors but adjacent filaments can have strongly deviant motions.  The influence of voids must be very important.  Void evacuation patterns are becoming increasingly apparent.  Filaments are walls of these voids and adjacent filaments can be responding in their own way to void geometries.

\bigskip\bigskip
This project has been going on for a long time and there are a lot of people to thank.  There are the young people who have helped out on their ways to other adventures: Austin Barnes, Nicolas Bonhomme, Emily Chang, Bryson Yee, Matt Zagursky, Max Zavodny.  So much of our products build on the efforts of others: the late Tony Fairall, Stephane Courteau, Riccardo Giovanelli, Martha Haynes, Ren\'ee Kraan-Korteweg, Karen Masters, Jeremy Mould, Chris Springob, Kartik Sheth and our sometimes collaborators Kristin Chiboucas, Renzo Sancisi, Will Saunders, and Marc Verheijen.  Our enterprise would be much the poorer without our theory collaborators Stefan Gottl\"ober, Yehuda Hoffman, Jim Peebles, and Stephen Phelps, or our visualization specialist Daniel Pomar\`ede.  We are in the midst of new opportunities for photometry with Spitzer and WISE space telescopes with collaborators Tom Jarrett, Don Neill, Mark Seibert, Wendy Freedman, and Barry Madore.  Mentioning Barry, we are reminded of the tremendous importance of NED, the NASA-IPAC Extragalactic Database, and also of LEDA, the Lyon Extragalactic Database and Georges Paturel and Philippe Prugniel.  Several times every day we look at one or other of these resources.  We thank James Shombert for his development of the Archangel photometry package and Stuart Levy for his development of the Partiview visualization program.  Finally, where would any study of large scale structure be without the immense contribution of our dear departed friend John Huchra?
Funding support has come from the US National Science Foundation award AST-0908846, the NASA ADAP award NNX12AE70G, the Spitzer Space Telescope cycle 8 program 80072, and HST programs GO-9162, 10210, 10905, 11285, 11584, 12546, and 12878.
I. Karachentsev, D. Makarov and S. Mitronova acknowledge support by the RFBR grant no. 11-02-00639 and the grant of the Ministry of Education and Science of the Russian Federation no. 8523.

\clearpage

\begin{table}[htbp]
  \center
  \caption{Cosmicflows-2 Compendium of Distances}
    \begin{tabular}{rrccccccccccccrrrrrcrrrrrrrlrrrrccrrrrrrrrrrrrrrr}
\hline
    PGC &   d    &  dm   &  ed  & C & T & L & M & S & N & H & F &    RAJ   &    DeJ    &    l     &    b     &   sgl    &   sgb    &  T &  Ag  &  Bt   &  Ks   &  Vhel &  Vgsr &   Vls &  Vcmb &  Vmod &    Name    & GpID & 2M++ &  Nd &   dg   &  dmg  & edg  &  NV &   lg     &   bg     &  sglg    &  sgbg    &Tg&$\Sigma$B&$\Sigma$K&Vhelg& Vgsrg &  Vlsg & Vcmbg & Vmodg & sigV & Group Name     \\
\hline
      4 &  50.58 & 33.52 & 0.20 &   &   &   &   &   &   & H &   & 000003.5 & +230515.5 & 107.8322 & -38.2729 & 316.0587 &  18.4514 &  5 & 0.40 & 16.88 &  0.00 &  4458 &  4638 &  4706 &  4109 &  4154 &  AGC331060 &      &      &   1 &  50.58 & 33.52 & 0.20 &   1 & 107.8322 & -38.2729 & 316.0587 &  18.4514 &  5 & 16.88 &  0.00 &  4458 &  4638 &  4706 &  4109 &  4154 &    0 &                  \\       
     55 &  73.79 & 34.34 & 0.20 &   &   &   &   &   &   & H &   & 000037.4 & +333603.4 & 110.9496 & -28.0856 & 327.0998 &  19.7763 &  6 & 0.22 & 17.04 &  0.00 &  4779 &  4979 &  5052 &  4454 &  4507 &   UGC12898 &      &      &   1 &  73.79 & 34.34 & 0.20 &   1 & 110.9496 & -28.0856 & 327.0998 &  19.7763 &  6 & 17.04 &  0.00 &  4779 &  4979 &  5052 &  4454 &  4507 &    0 &                    \\     
     70 & 117.49 & 35.35 & 0.20 &   &   &   &   &   &   & H &   & 000056.1 & +202016.7 & 107.1780 & -40.9837 & 313.2487 &  17.7662 &  6 & 0.34 & 15.61 & 11.26 &  6800 &  6974 &  7040 &  6447 &  6557 &   UGC12900 &      &      &   1 & 117.49 & 35.35 & 0.20 &   1 & 107.1780 & -40.9837 & 313.2487 &  17.7662 &  6 & 15.61 & 11.26 &  6800 &  6974 &  7040 &  6447 &  6557 &    0 &                      \\   
     76 &  97.72 & 34.95 & 0.20 &   &   &   &   &   &   & H &   & 000058.9 & +285441.5 & 109.8059 & -32.6707 & 322.1728 &  19.1316 &  3 & 0.21 & 14.82 & 11.01 &  6920 &  7112 &  7183 &  6583 &  6697 &   UGC12901 &      &      &   1 &  97.72 & 34.95 & 0.20 &   1 & 109.8059 & -32.6707 & 322.1728 &  19.1316 &  3 & 14.82 & 11.01 &  6920 &  7112 &  7183 &  6583 &  6697 &    0 &                        \\ 
    124 &  81.66 & 34.56 & 0.20 &   &   &   &   &   &   & H &   & 000136.7 & +033020.1 &  99.5866 & -57.0904 & 296.2413 &  13.7943 &  6 & 0.10 & 16.90 &  0.00 &  6350 &  6477 &  6529 &  5988 &  6083 &   UGC12913 &      &      &   1 &  81.66 & 34.56 & 0.20 &   1 &  99.5866 & -57.0904 & 296.2413 &  13.7943 &  6 & 16.90 &  0.00 &  6350 &  6477 &  6529 &  5988 &  6083 &    0 &                         \\
    143 &   0.96 & 24.92 & 0.05 & C & T & L & M &   &   &   &   & 000158.2 & -152739.3 &  75.8635 & -73.6245 & 277.8077 &   8.0860 & 10 & 0.16 & 11.04 &  9.00 &  -108 &   -45 &   -14 &  -441 &  6083 &        WLM &  222 &      &  39 &   0.76 & 24.40 & 0.01 &  41 & 125.4790 & -26.4230 & 322.2989 &   9.4579 &  2 &  3.40 &  0.68 &  -213 &   -49 &    17 &  -485 &  6083 &  156 &    2557    NGC0224      \\
    145 & 134.28 & 35.64 & 0.20 &   &   &   &   &   &   & H &   & 000157.2 & -275957.9 &  25.5365 & -79.0034 & 265.9224 &   3.8952 &  3 & 0.08 & 15.06 &  0.00 &  9723 &  9739 &  9753 &  9427 &  9661 & ESO409-004 &      &      &   1 & 134.28 & 35.64 & 0.20 &   1 &  25.5365 & -79.0034 & 265.9224 &   3.8952 &  3 & 15.06 &  0.00 &  9723 &  9739 &  9753 &  9427 &  9661 &    0 &                         \\
    165 &  86.30 & 34.68 & 0.20 &   &   &   &   &   &   & H &   & 000223.0 & +271238.0 & 109.6981 & -34.3974 & 320.4400 &  18.5834 &  4 & 0.21 & 15.77 &  0.00 &  7620 &  7808 &  7878 &  7280 &  7420 &   UGC12920 &      &      &   1 &  86.30 & 34.68 & 0.20 &   1 & 109.6981 & -34.3974 & 320.4400 &  18.5834 &  4 & 15.77 &  0.00 &  7620 &  7808 &  7878 &  7280 &  7420 &    0 &                         \\
    176 & 103.28 & 35.07 & 0.20 &   &   &   &   &   &   & H &   & 000234.8 & -034238.9 &  94.3256 & -63.8349 & 289.2040 &  11.5576 &  4 & 0.17 & 14.49 & 10.98 &  6465 &  6568 &  6612 &  6109 &  6208 &  AGC400001 &      &      &   1 & 103.28 & 35.07 & 0.20 &   1 &  94.3256 & -63.8349 & 289.2040 &  11.5576 &  4 & 14.49 & 10.98 &  6465 &  6568 &  6612 &  6109 &  6208 &    0 &                         \\
    179 &  71.12 & 34.26 & 0.20 &   &   &   &   &   &   & H &   & 000239.9 & +084413.0 & 103.0012 & -52.2368 & 301.5334 &  14.8675 &  3 & 0.28 & 15.69 & 11.48 &  5537 &  5680 &  5736 &  5175 &  5246 &  AGC100002 &      &      &   1 &  71.12 & 34.26 & 0.20 &   1 & 103.0012 & -52.2368 & 301.5334 &  14.8675 &  3 & 15.69 & 11.48 &  5537 &  5680 &  5736 &  5175 &  5246 &    0 &                         \\
    186 & 115.88 & 35.32 & 0.20 &   &   &   &   &   &   & H &   & 000246.4 & +185310.6 & 107.2408 & -42.4978 & 311.8408 &  17.0661 &  1 & 0.15 & 14.82 & 10.91 &  7894 &  8063 &  8128 &  7539 &  7689 &   UGC00003 &      &      &   1 & 115.88 & 35.32 & 0.20 &   1 & 107.2408 & -42.4978 & 311.8408 &  17.0661 &  1 & 14.82 & 10.91 &  7894 &  8063 &  8128 &  7539 &  7689 &    0 &                         \\
    201 & 127.06 & 35.52 & 0.20 &   &   &   &   &   &   & H &   & 000257.0 & +041231.0 & 100.6048 & -56.5458 & 297.0307 &  13.6568 &  4 & 0.12 & 15.52 & 11.71 &  8629 &  8758 &  8810 &  8267 &  8447 &   UGC00004 &      &      &   1 & 127.06 & 35.52 & 0.20 &   1 & 100.6048 & -56.5458 & 297.0307 &  13.6568 &  4 & 15.52 & 11.71 &  8629 &  8758 &  8810 &  8267 &  8447 &    0 &                         \\
    205 & 105.20 & 35.11 & 0.20 &   &   &   &   &   &   & H &   & 000305.7 & -015449.9 &  96.2231 & -62.2521 & 290.9996 &  11.9504 &  4 & 0.17 & 14.17 & 10.42 &  7267 &  7376 &  7422 &  6910 &  7036 &   UGC00005 &    0 & 3703 &   1 & 105.20 & 35.11 & 0.20 &   3 &  96.9948 & -62.1307 & 291.3296 &  11.7497 &  4 &       & 10.05 &  7218 &  7328 &  7374 &  6861 &  6985 &   45 &                UGC00005 \\
    212 & 206.06 & 36.57 & 0.20 &   &   &   &   &   &   & H &   & 000311.2 & +155755.6 & 106.3213 & -45.3440 & 308.8750 &  16.3795 &  2 & 0.19 & 14.94 & 10.65 & 11229 & 11391 & 11454 & 10871 & 11182 &   UGC00007 &      &      &   1 & 206.06 & 36.57 & 0.20 &   1 & 106.3213 & -45.3440 & 308.8750 &  16.3795 &  2 & 14.94 & 10.65 & 11229 & 11391 & 11454 & 10871 & 11182 &    0 &                         \\
    218 &  13.68 & 30.68 & 0.12 &   &   &   &   & S &   & H &   & 000314.9 & +160844.0 & 106.4096 & -45.1747 & 309.0615 &  16.4021 &  2 & 0.19 & 11.59 &  7.18 &  1051 &  1213 &  1276 &   694 &   695 &   NGC07814 & 1211 &      &   3 &  12.30 & 30.45 & 0.09 &   4 & 107.6898 & -45.4946 & 309.1776 &  15.4437 &  8 & 11.08 &  7.13 &   911 &  1071 &  1133 &   555 &   556 &   96 &     218   NGC07814      \\
    226 &  92.47 & 34.83 & 0.20 &   &   &   &   &   &   & H &   & 000320.4 & +083708.0 & 103.2027 & -52.4005 & 301.4568 &  14.6765 &  5 & 0.32 & 15.36 & 11.26 & 11946 & 12088 & 12144 & 11584 & 11937 &   UGC00010 &      &      &   1 &  92.47 & 34.83 & 0.20 &   1 & 103.2027 & -52.4005 & 301.4568 &  14.6765 &  5 & 15.36 & 11.26 & 11946 & 12088 & 12144 & 11584 & 11937 &    0 &                         \\
    250 &  92.04 & 34.82 & 0.08 &   &   &   &   &   & N & H &   & 000335.0 & +231202.9 & 108.8693 & -38.3606 & 316.3211 &  17.6720 &  6 & 0.41 & 14.42 & 10.95 &  7265 &  7444 &  7512 &  6917 &  7043 &   UGC00014 &      &      &   1 &  92.04 & 34.82 & 0.08 &   1 & 108.8693 & -38.3606 & 316.3211 &  17.6720 &  6 & 14.42 & 10.95 &  7265 &  7444 &  7512 &  6917 &  7043 &    0 &                         \\
    255 &  20.32 & 31.54 & 0.20 &   &   &   &   &   &   & H &   & 000343.2 & +151305.4 & 106.2143 & -46.0988 & 308.1421 &  16.0963 &  9 & 0.21 & 15.09 &  0.00 &   878 &  1038 &  1100 &   520 &   521 &   UGC00017 & 1211 &      &   3 &  12.53 & 30.49 & 0.09 &   4 & 107.6898 & -45.4946 & 309.1776 &  15.4437 &  8 & 11.08 &  7.13 &   911 &  1071 &  1133 &   555 &   556 &   96 &     218   NGC07814      \\
    259 & 109.14 & 35.19 & 0.20 &   &   &   &   &   &   & H &   & 000344.3 & +161112.4 & 106.5852 & -45.1656 & 309.1290 &  16.2962 &  4 & 0.20 & 15.67 &  0.00 &  6412 &  6574 &  6637 &  6055 &  6152 &  AGC100020 &      &      &   1 & 109.14 & 35.19 & 0.20 &   1 & 106.5852 & -45.1656 & 309.1290 &  16.2962 &  4 & 15.67 &  0.00 &  6412 &  6574 &  6637 &  6055 &  6152 &    0 &                         \\
    265 & 118.58 & 35.37 & 0.20 &   &   &   &   &   &   & H &   & 000351.7 & -504559.8 & 320.6851 & -64.6904 & 244.6595 &  -4.2980 &  4 & 0.06 & 16.13 &  0.00 & 11900 & 11829 & 11813 & 11708 & 12068 &  PGC000265 &      &      &   1 & 118.58 & 35.37 & 0.20 &   1 & 320.6851 & -64.6904 & 244.6595 &  -4.2980 &  4 & 16.13 &  0.00 & 11900 & 11829 & 11813 & 11708 & 12068 &    0 &                         \\
\hline
    \end{tabular}%
  \label{tab:alld}%
\end{table}%

\begin{table}[htbp]
  \center
  \caption{Distances for Galaxy Groups}
    \begin{tabular}{rrrrrrrrrrrrrrrrrrrrrrrr}
    \hline
   Nd &    dg  &   dmg  &   edg &   NV &     lg    &     bg    &    sglg   &    sgbg   &  Tg&$\Sigma$B&$\Sigma$K&Vhelg&  Vgsrg &   Vlsg & Vcmbg  & Vmodg  & sigV  &   Vpec &  cVpec & icnt & 2m++ &   PGC      Name\\
\hline
   16 &   0.01 &  14.50 &  0.02 &   20 &    0.0000 &    0.0000 &  185.7861 &   42.3103 &   1 &  -6.   & -9.    &   51 &     26 &     13 &     91 &     91 &    64 &      0 &      0 &  223 &    0 &       0     Galaxy\\
   39 &   0.76 &  24.40 &  0.01 &   41 &  125.4790 &  -26.4230 &  322.2989 &    9.4579 &   2 &   3.40 &  0.68  & -213 &    -49 &     17 &   -485 &   -485 &   156 &      0 &      0 &  222 &    0 &    2557    NGC0224\\
    4 &   1.37 &  25.68 &  0.03 &    4 &  251.1365 &   32.2601 &  120.5020 &  -42.5472 &  10 &   9.75 &  9.29  &  347 &    153 &    103 &    690 &    691 &    23 &    589 &    589 &  227 &    0 &   29128  NGC003109\\
    1 &   1.37 &  25.68 &  0.10 &    1 &  164.6636 &   42.8855 &   47.6118 &  -15.0125 &  10 &  13.75 & 11.50  &  -29 &     17 &     45 &    130 &    130 &     0 &     28 &     28 &    0 &    0 &   26142    UGC4879\\
    1 &   1.91 &  26.41 &  0.08 &    1 &   83.8788 &   44.4092 &   56.0935 &   40.3700 &  10 &  16.69 & 12.90  & -139 &     44 &     78 &   -121 &   -121 &     0 &      0 &      0 &  230 &    0 & 2801026      KKR25\\
    7 &   2.02 &  26.53 &  0.03 &    7 &  240.5522 &  -70.3721 &  254.5662 &    3.1429 &   4 &   7.64 &  5.94  &   90 &     72 &     71 &   -158 &   -158 &    30 &      0 &      0 &  234 &    0 &    1014  NGC000055\\
    4 &   2.26 &  26.77 &  0.04 &    4 &  184.7665 &   70.5101 &  100.4254 &   17.0085 &  10 &  13.67 & 11.70  &  179 &    194 &    172 &    427 &    427 &    34 &    259 &    259 &  229 &    0 &   50961    UGC9128\\
    1 &   2.55 &  27.03 &  0.10 &    1 &  328.5515 &  -17.8494 &  199.1885 &    8.6107 &  10 &  11.52 &  9.45  &  306 &    188 &    130 &    323 &    323 &     0 &    133 &    133 &  244 &    0 &   60849     IC4662\\
    1 &   2.64 &  27.11 &  0.08 &    1 &  111.1420 &   61.3082 &   63.0900 &   17.9058 &  10 &  13.94 & 11.59  &   59 &    176 &    198 &    196 &    196 &     0 &      0 &      0 &  215 &    0 &   47495    UGC8508\\
    2 &   2.80 &  27.24 &  0.05 &    2 &  118.0639 &  -24.6938 &  332.1004 &   14.7111 &   4 &  10.93 &  7.61  &  -90 &    101 &    175 &   -388 &   -388 &    30 &      0 &      0 &  224 &    0 &    4126     NGC404\\
    2 &   2.86 &  27.28 &  0.07 &    2 &  302.0653 &  -15.8315 &  194.6908 &  -16.3986 &  10 &  11.71 &  9.12  &  421 &    221 &    152 &    533 &    534 &    12 &    321 &    321 &  252 &    0 &   39573     IC3104\\
    7 &   2.87 &  27.29 &  0.03 &    7 &  158.0996 &   74.7942 &   75.9126 &    0.9943 &  10 &   9.88 &  7.88  &  223 &    252 &    252 &    471 &    472 &    55 &    258 &    258 &  214 &    0 &   39225    NGC4214\\
    4 &   3.10 &  27.45 &  0.04 &    4 &   79.8128 &   71.2497 &   79.2071 &   21.7476 &  10 &  12.65 & 10.02  &  192 &    278 &    282 &    375 &    375 &    11 &    144 &    144 &  213 &    0 &   51472     DDO190\\
   10 &   3.18 &  27.51 &  0.03 &   19 &  137.6285 &    6.7613 &  120.4475 &    0.4828 &   8 &   6.01 &  4.26  &   32 &    190 &    258 &    -78 &    -78 &    85 &      0 &      0 &  221 &    0 &   13826     IC0342\\
    1 &   3.55 &  27.75 &  0.10 &    1 &   94.9743 &   21.5182 &   23.2744 &   41.5918 &  10 &  13.55 & 11.17  & -139 &     94 &    153 &   -262 &   -262 &     0 &      0 &      0 &  228 &    0 &   63000    NGC6789\\
    9 &   3.61 &  27.79 &  0.03 &    9 &  110.1049 &  -81.3626 &  270.2023 &   -1.2171 &   7 &   7.48 &  3.75  &  207 &    218 &    235 &    -81 &    -81 &    85 &      0 &      0 &  233 &    0 &    2789  AGC020535\\
   45 &   3.66 &  27.82 &  0.01 &   47 &  142.5322 &   39.7347 &   40.0558 &    0.0403 &   7 &   6.65 &  3.28  &   40 &    154 &    198 &    117 &    117 &   106 &   -155 &   -155 &  217 &    0 &   28630    NGC3031\\
   26 &   3.68 &  27.83 &  0.02 &   27 &  309.9711 &   19.4277 &  159.7460 &   -4.8335 &   3 &   6.80 &  3.31  &  350 &    179 &    105 &    603 &    604 &   283 &    330 &    330 &  240 &    0 &   46957    NGC5128\\
    2 &   3.95 &  27.99 &  0.07 &    2 &  273.3902 &  -71.7079 &  256.2020 &  -18.7344 &  10 &  11.25 &  8.95  &  395 &    310 &    302 &    209 &    209 &    11 &    -85 &    -82 &  235 &    0 &    5896  NGC000625\\
    3 &   4.02 &  28.02 &  0.06 &    5 &  322.6940 &   11.1363 &  169.4448 &    6.4211 &   9 &  10.47 &  7.98  &  552 &    413 &    342 &    731 &    732 &   101 &    433 &    433 &  242 &    0 &   54392  ES274-001\\
\hline
    \end{tabular}%
  \label{tab:groupd}%
\end{table}%



\clearpage

\bibliographystyle{Apj}

\bibliography{bibli2}

\end{document}